\documentclass[referee]{aa}
\usepackage[varg]{txfonts}
\usepackage{graphicx}
\usepackage{natbib}

\def\figfigincl#1#2#3{\includegraphics[width=#1]{figures/#2.eps}%
    \caption{\small #3}\label{fig:#2}}
\def\figfiginclrot#1#2#3{\includegraphics[width=#1,angle=-90]{figures/#2.eps}%
    \caption{\small #3}\label{fig:#2}}
\def\figfiginclnocap#1#2{\includegraphics[width=#1]{figures/#2.eps}}
\def\figfiginclrotnocap#1#2{\includegraphics[width=#1,angle=-90]{figures/#2.eps}}

\begin{document}

\title{Asteroid cratering families: recognition and collisional interpretation} 

\author{A. Milani\inst{1}\thanks{\emph{In the course of the final revision of this paper, Andrea Milani suddenly passed away. We want to dedicate this paper to his memory. Andrea, you will be dearly missed.}}
        \and Z. Kne\v zevi\'c\inst{2}
        \and F. Spoto\inst{3} 
        \and P. Paolicchi\inst{4}
   }

\institute{Dept.Mathematics, University of Pisa, Largo Pontecorvo 5,
  I-56127 Pisa, Italy\\ \email{milani@dm.unipi.it}
  \and Serbian Academy Sci. Arts, Kneza Mihaila 35, 11000 Belgrade,
  Serbia \\\email{zoran.knezevic@sanu.ac.rs} 
  \and IMCCE, Observatoire de Paris, PSL Research University, CNRS,
  Sorbonne Universités, UPMC Univ. Paris 06, Univ. Lille 77
  av. Denfert-Rochereau, 75014, Paris, France \\ \email{fspoto@oca.eu}
  \and Dept.Physics, University of Pisa, Largo Pontecorvo 3, I-56127
  Pisa, Italy\\ \email{paolo.paolicchi@unipi.it}}

\date{19 November 2018}


\abstract {} {We continue our investigation of the bulk properties of
  asteroid dynamical families identified using only asteroid proper
  elements \citep{bigfamilies} to provide plausible collisional
  interpretations. We focus on cratering families consisting of a
  substantial parent body and many small fragments.}  {We propose a
  quantitative definition of cratering families based on the fraction
  in volume of the fragments with respect to the parent body;
  fragmentation families are above this empirical boundary.  We assess
  the compositional homogeneity of the families and their shape in
  proper element space by computing the differences of the proper
  elements of the fragments with respect to the ones of the major
  body, looking for anomalous asymmetries produced either by
  post-formation dynamical evolution, or by multiple
  collisional/cratering events, or by a failure of the Hierarchical
  Clustering Method (HCM) for family identification.} {We identified a
  total of 25 dynamical families with more than $100$ members ranging
  from moderate to heavy cratering.  For three families (4, 15 and
  283) we confirm the occurrence of two separate cratering events,
  while family (569) Misa is a mixed case, with one cratering event
  and one fragmentation event.  The case of family 3 remains dubious,
  in that there could be either one or two collisions.  For family 20,
  we propose a double collision origin, not previously identified.  In
  four cases (31, 480, 163 and 179) we performed a dedicated search
  for dynamical resonant transport mechanisms that could have
  substantially changed the shape of the family. By using a new
  synthetic method for computation of secular frequencies, we found
  possible solutions for families 31, 480, and 163, but not for family
  179, for which we propose a new interpretation, based on a secular
  resonance contaminating this family: the family of 179 should be
  split into two separate clusters, one containing (179) itself and
  the other, family (9506) Telramund, of fragmentation type, for which
  we have computed an age.}{}


\maketitle

\section{Introduction}
\label{s:intro}

The method that we proposed to study asteroid families
\citep{bigfamilies} and have applied in our research over the last few years
\citep{paperII, paperIII, paperIV, paperV} is as follows. We first
compute (and periodically update) very large catalogs of asteroid
proper elements, using the synthetic method
\citep{knemil2000}. We then use a modified form of the Hierarchical
Clustering Method (HCM) \citep{zappala90} to identify clusters,
representing density contrast in the three-dimensional (3D) space of proper
elements $a, e, \sin{I}$. After confirming some of the clusters with a
statistical significance test, we proclaim these as \textit{dynamical
  families}. The classification is updated when the catalogs of proper
elements are increased.

In the second stage we add physical observation data (absolute
magnitude and albedo) to estimate ages of these families by exploiting
nongravitational perturbations, which are size dependent. This
sometimes forces the splitting of dynamical families into multiple
\textit{collisional families} of different ages.  In the third stage,
specific to this paper, we use available physical data remove
interlopers, that is background asteroids randomly present in the
volume occupied by the families.

This complex procedure is used because the amount of
information contained in the proper elements is greater (by a factor of
at least $5$; see Table~1 in \citet{bigfamilies}) than the information
contained in consistent catalogs of physical observations:
the proper element sets are both more accurate and more
numerous. Because families are statistical entities, the number of
data records matters more than anything else. In particular, the
proper element catalogs contain much smaller asteroids than the ones
for which good physical observations are available, for obvious
observational selection effects: this is especially important in the
context of the present paper, which is about cratering families, that
typically have smaller members.

As a result of this procedure, we believe we can claim that our work
is mathematically sound in all steps, including the computation of
proper elements, the application of HCM, and the computation of ages.
However, the use of rigorous mathematics is a necessary condition for
the successful application of a mathematical model in science, not a
sufficient one. We need to test whether the use of our sophisticated
model actually leads to a better understanding of the collisional
history of the asteroid belts. The question to be addressed is whether
our methods produce superior results to the ones obtained with simpler
models: lower-accuracy proper elements (e.g., analytical rather than
synthetic), and simpler clustering methods (e.g., the ones not
including a test of the statistical significance of the families).

The main goal of this paper is to display a large set of data,
summarized in tables, describing detailed properties of the
collisional events giving rise to the cratering families (in full or,
in some cases, in part). If these data are meaningful, then not just
the classification, which is the existence of the most important
families (e.g., the ones with $>100$ members) and the decomposition of
some of them in multiple collisional families, but even details of the
shape of the families in the space of proper elements contains
information constraining the collisional event itself. Although we
cannot claim that all our families have ``good shapes'', because some
problematic cases still need more data and more work, the fact is that
the large majority of them are compatible with simple and physically
possible collisional models, for example, with ejection velocities of
the same order of magnitude as the escape velocity from the parent
body.

Although the arguments above can apply to the outcomes of all
collisional events, those pertaining to cratering type, on which we
are focusing in this paper, are especially interesting for three
reasons: first, by forming families with smaller members, we have
found many more cratering families than previously known.
Second, by definition (see Section~\ref{s:recrat}), a cratering event
leaves a parent body with essentially the same collisional cross
section as before, and therefore it is normal that another crater can be
formed, leading to overlapping families.
Third, a cratering event must generate a distribution of relative
velocities strongly anisotropic with respect to the parent body, and
this should be detectable from the family shape. In some cases we have
not found the expected anisotropy, but either an overly small one,
which can suggest a multiple collisional origin, or an overly large one,
corresponding to unrealistic escape velocities, indicating a
post-collision dynamical evolution.

Obtaining such a well behaved original escape velocity distribution is
by no means guaranteed. There are necessarily cases in which the
distribution is strange, even bizarre, and these cases are analyzed to
find an explanation. Two possible causes for such strange
distributions are already clear, namely the possibility of detecting a
multiple collisional origin that was not previously recognized, and
the presence of a dynamical mechanism, such as resonances and chaos,
leading to an important change in the shape of the family after its
formation. A third possibility, which has been largely overlooked so
far, is that statistical methods, HCM not excluded, cannot provide
absolute certainty on the membership of families; in some critical
cases not even on their statistical significance. The presence of
interlopers in any family is a well known phenomenon that needs to be
accounted for as well as possible \citep{radonov}, but in few cases
substantial corrections to the family membership may be needed,
including merging and splitting of previously identified families. In
this paper we deal with examples of both merge and split corrections.

The paper is organized as follows: in Section~\ref{s:recrat} we
discuss the quantitative definition of cratering and present our
proposed list of cratering families. The data include the fraction (by
volume) of fragments with respect to the total reconstructed volume of
the parent body, and the data on interlopers which could be
recognized. In Section~\ref{s:properties} we measure the asymmetry of
the shape of the family in proper element space, and discuss its
interpretation. In Section~\ref{s:shapes} we discuss possible
dynamical, statistical and/or collisional interpretation of family
shapes which do not have a simple interpretation in terms of original
escape velocities.  In Section~\ref{s:conclusions} we draw our
conclusions and outline possible future work. In Appendix A we
present an empirical model showing that the initial change in the proper
semimajor axis (with respect to the parent body) can be either
correlated or anticorrelated with the Yarkovsky secular effect.  In
Appendix~B we present a compilation of the data on the age of the
collisional events in the cratering families, which are mostly not new
results but collected for the convenience of the reader from our
previous papers \citep{paperIII,paperIV,paperV}; two new ages have
been computed for the families of (87) Sylvia and (9506) Telramund.

\section{Recognition of cratering families}
\label{s:recrat}

An asteroid family is of \textit{cratering type} if it is formed by a
collision leaving a parent body with the same impact cross section
and a new, large crater on its surface.  It is of
\textit{fragmentation type} if the largest fragment is significantly
smaller than the parent body.

In practice, it is comparatively easy to label some families as
cratering type, especially when concerning the largest asteroids (like
(4) Vesta, (10) Hygiea, and (2) Pallas) because of a huge gap in size
between the namesake asteroid and the other family members.

It is also easy to label some families as fragmentation type, like the
one of (24) Themis, because it can be estimated that the parent body
was approximately of diameter $D=284$ km against $D=202$ km of (24)
Themis itself; that is, the parent body was significantly larger. An
example of total fragmentation is the family of (158) Koronis, named
after an asteroid which is not even the largest remnant but just the
one with the lowest number. (158) Koronis itself has $D=48$ km and the
parent body can be estimated to have had $D=96$ km.

There are, of course, intermediate cases in which the recognition of
cratering vs. fragmentation families is not obvious. 

Historically, at the beginning of the research on asteroid families,
no cratering families were known, because the first identified
families consisted only of comparatively large members
\citep{hirayama, brouwer}. Later, cratering families began to appear
because of the discovery of smaller and smaller asteroids. The most
striking example is the family of (4) Vesta which appeared as a
``tiny'' family (7 members) in \citet{zappala90}, then as a ``small''
family (64 members) in \citet{zappala95}. Later \citet{knemil2003}
argued that the Vesta family should have thousands of members, almost
all smaller than $D=7$ km, with a very large spread in proper $a$,
more than $0.1$ au, although the numbers were so large that the
methods of classification available at the time were not
adequate. Indeed, with the multistep HCM classification method of
\citet{bigfamilies}, the Vesta family was found to have $7,865$
members, and in the latest classification update is seen to have
$10,612$ members, with a spread in proper $a$ of $0.246$ au. Both
properties are now fully understood: the small size of fragments,
because of the limit of the size of fragments from a cratering event
(controlled by the depth, not the diameter of the crater), and the
spread in proper $a$ because of the Yarkovsky effect (changing proper
$a$ in a way proportional to the time elapsed from the collision and
inversely proportional to the fragment diameter).

Now, with our most recent classification including $130,933$ members
in $118$ families, we can find many cratering families not previously
recognized as such. In light of these new findings however a clear
criterion for cratering families needs to be defined.

We propose a method to identify cratering families by using the
fraction (in volume) of fragments with respect to the total (including
the parent body). This fraction needs to be computed after removing
from the membership list both \textit{interlopers} (identified by
means of physical observations showing incompatible composition) and
\textit{outliers} (excluded from the V-shape fit used to estimate the
age; see \citet{paperIII}).  These removals may significantly affect
the fraction of fragments, because the family has a steeper size
distribution with respect to the background.

\subsection{List of cratering families}
\label{s:list}

\begin{table}[t!]
\centering
\caption{Cratering families: dynamical family number, collisional
  family number, number of members with high albedo, number of members
  with low albedo, diameter of the largest member, total number of
  members, fraction of fragments by volume, family albedo,
  taxonomy of the largest member, and notes.}

\medskip
\begin{tabular}{|l|l|rr|r|r|c|l|c|l|}

\hline
ND  &  NC & brig& dark& D   &  ntot &  $f_v$  &  malb & Cl &note\\ 
\hline
3   &  3  &  79 &  13 & 247 &  1693 & \textbf{0.0007} &  0.238& S &2? ag\\ %
4   &  4  & 802 &  32 & 525 & 10610 & \textbf{0.0003} &  0.423& V &2 ag\\  %
5   &  5  & 334 & 186 & 115 &  6169 & \textbf{0.0502} &  0.245& S &   \\   %
10  & 10  &  45 & 852 & 453 &  3143 & \textbf{0.0231} &  0.073& C &   \\  
15  & 15  &2230 &  92 & 232 &  9854 & \textbf{0.1160} &  0.260& S &2 ag\\  %
20  & 20  & 142 &  13 & 136 &  7818 & \textbf{0.0072} &  0.210& S &\\      %
31  & 31  &   0 & 555 & 267 &  1385 & \textbf{0.0274} &  0.045& C &\\    %
87  & 87  &   0 &  75 & 288 &   191 & \textbf{0.0080} &  0.046& C &OS\\   
163 & 163 &  10 & 311 &  82 &  1021 & \textbf{0.1288} &  0.033& C &merg\\  %
283 & 283 &   2 & 199 & 135 &   577 & \textbf{0.0807} &  0.032& C &2 ag\\  %
569 & 569 &   2 & 159 &  73 &   646 & \textbf{0.1221} &  0.030& C &sub\\   %
194 & 686 &  43 &  12 &  55 &   376 & \textbf{0.0735} &  0.142& I &NC\\ %
\hline 
302 & 302 &   0 &  42 &  39 &   233 & \textbf{0.0548} &  0.052& C &Y\\     %
396 & 396 &   6 &   7 &  37 &   529 & \textbf{0.0560} &  0.139& I &Y\\     %
606 & 606 &   5 &   8 &  37 &   325 & \textbf{0.1230} &  0.089& I &Y\\     %
\hline
110 & 363 &   4 & 348 &  97 &   894 & \textbf{0.1723} &  0.059& C &NC\\    %
480 & 480 & 229 &   1 &  56 &  1162 & \textbf{0.1575} &  0.249& S &\\    
1303&1303 &   1 & 114 &  92 &   232 & \textbf{0.1407} &  0.052& C &\\    %
1547& 1547&  11 &   1 &  21 &   344 & \textbf{0.2108} &  0.203& S &R?\\    
\hline
2   & 2   &  11 &   5 & 544 &    62 & \textbf{0.0003} &  0.101& I &sat\\%
96  & 96  &   0 &  61 & 177 &   120 & \textbf{0.0108} &  0.048& C &NA\\    %
148 & 148 &  18 &   0 &  98 &   137 & \textbf{0.0009} &  0.164& S &NA\\    %
179 & 179 &  54 &   9 &  73 &   513 & \textbf{0.0469} &  0.198& S &NA\\    %
410 & 410 &   3 &  32 & 118 &   120 & \textbf{0.0764} &  0.043& C &NA\\
778 & 778 &   0 &  86 & 118 &   574 & \textbf{0.0373} &  0.079& C &R\\    %
1222&1222 &   5 &   4 &  29 &   107 & \textbf{0.0872} &  0.165& S &NA\\
\hline
\end{tabular}

\label{tab:frac}
\end{table}

The results of the analysis of the $66$ families with more than $100$
members in our current updated classification\footnote{Available from
  AstDyS at \\http://hamilton.dm.unipi.it/astdys/index.php?pc=5} are
summarized in Table~\ref{tab:frac}. The horizontal lines split into
cratering with age, young ($<100$ Myr of age) cratering, heavy
cratering $f_v>0.125$, and cratering without age.

Included in the table are all the cases where the fraction in volume
of the fragments (with respect to the estimated volume of the parent
body) is $f_v<0.22$; this could be a critical value for the definition
of cratering. The adopted boundary may appear somewhat arbitrary, but
the fact is that we have found a gap in the values, that is, there are
no families with $0.22<f_v<0.26$. Hence, we can define families of
\textbf{cratering} type those with $f_v<1/4$ and families of
\textbf{fragmentation} type as those with $f_v>1/4$. Above this value
we have a region with few $f_v$ values, extending up to $0.50$. Due to
the lack of an explicit model justifying the choice of a very specific
value for this boundary, we propose also to define \textbf{heavy
  cratering} families as those with $1/8<f_v<1/4$ and \textbf{marginal
  fragmentation} families as those with $1/4<f_v<1/2$. A family with
$1/8<f_v<1/4$ still has most of the mass, and therefore most of the
internal structure, in the parent body and the impact cross section is
not significantly changed; however we should expect that most of the
original surface has not been preserved.

In summary, we propose a list of $15$ cratering families for which
we have computed (in one of our previous papers) an age, $4$ heavy
cratering families (also with age), and $6$ cratering families for
which we have no age estimate. Moreover there is the already mentioned
family of (2) Pallas which can be recognized as cratering even though
it has only $62$ members (even after merging $45$ members of the
family of (2) with the $17$ members of the family of (14916) 1993
VV$_7$, because they are separated by the three body resonance
$3J-1S-1A$).

The family of (87) Sylvia is strongly depleted by mean motion
resonances, especially $9/5$ and $11/6$ with Jupiter, and the low $a$
side appears to be missing entirely. Nevertheless it must belong to
the cratering type, although the current $f_v<0.01$ might have
originally been larger by a factor of several. The age of this family
has been computed in this paper at $1120\pm 282$ Myr; see
Appendix~\ref{s:age}.

The cratering families for which we have not computed an age include
those of (179) Klytaemnestra (Section~\ref{s:shapes}) and of (778)
Theobalda; an age of $6.9\pm 2.3$ Myr was computed for the latter using
with two methods specific to very young families
\citep{novakovic2010}. The other four families have $\leq 140$ members
and are not yet suitable for a reliable application of our V-shape
method. 

The four families we rate as marginal fragmentations are (808) Merxia
($f_v=0.2669$), (845) Naema ($0.3045$), (1118) Hanskya ($0.3709$) and
(1128) Astrid ($0.4020$). All the other cases tested ($37$ families
with $>100$ members) have $f_v>0.5$, and are therefore to be
considered of (nonmarginal) fragmentation type.

Table~\ref{tab:frac} is sorted by column 2, that is, by collisional
family number. This is due to the possibility of Name Change (note:
NC) when the namesake of the dynamical family is shown to be an
interloper, as for families 194 and 110. In this case the new name is
given by the lowest numbered member which is not an interloper: then
the diameter, the fraction of fragments, the albedo, and the taxonomy
refer to the new namesake.

Other notes in Tables~\ref{tab:frac} and \ref{tab:anisotr} have
the following meaning: 2 ag= 2 distinct ages; NA=no age computed; HI=
high proper inclination $>17^\circ$; OS=one sided V-shape;
sat=included satellite family; sub=with subfamily; merg=two families
merged because they form a single V-shape; Y=young families (age
$<100$ Myr); R=too recent ($< 10$ Myr) for V-shape computation of age.

In Table~\ref{tab:frac} the family albedo is the albedo of the parent
body, if available, otherwise the mean albedo of the members with
albedo measurements.  The taxonomy is given by large groups C, S, and
V; I corresponds to intermediate albedo.

The most problematic result of Table~\ref{tab:frac} concerns the
family of (569) Misa, because it contains a subfamily with namesake
(15124) 2000 EZ$_{39}$, and a much younger age (see
Table~\ref{tab:age_crat}), but the identity of its parent body is
unclear.  If it were (569), then it would be a second cratering, and
it should have analogous shape properties to the other craterings; see
Section~\ref{s:anisotropy}.

\subsection{Compositional homogeneity}

Columns 3, 4 and 9 of Table~\ref{tab:frac} contain data that can be
used to assess the homogeneity of the composition of the family. In
most cases the families with a parent body belonging to the C complex
of dark asteroids have an overwhelming majority of members with WISE
data \citep{WISE} showing a low albedo, see for examples families 10,
31, 163, 283, 569, and 363. Families with a parent body belonging to
the S complex of brighter asteroids, such as families 3, 15, 20, and
480, have a majority of bright members (the same for the V-type family
4), although the ratio bright/dark is not particularly
large\footnote{This is due to the observational selection typical of
  infrared instruments: there are many more low-albedo ($<0.1$)
  asteroids with WISE data than high albedo ($>0.2$) ones; the ones
  with albedo $>0.1$ are as many as those with $<0.1$, but this is due
  to a significant contribution from intermediate albedo between $0.1$
  and $0.2$.}

There are some cases where the composition appears more mixed, for
example families 5 and 686. For the dynamical family of prevalent type
S with parent body (5) Astraea, physical observations suggest the
presence of another family of type C asteroids locked in the same
secular resonance $g+g_5-2\, g_6$ as family 5, with parent body (91)
Aegina \citep{paperV}. Parent body (91) is large, with WISE $D=105$
km, and could not be a fragment of (5) due to both its size and its
composition. Even if all the other $185$ C-type interlopers in the
dynamical family 5 were members of family 91 (which is not necessarily
the case), family 5 would have $f_v=0.0239$, and would thereforse be
of cratering type nonetheless. Another interesting example of
cratering family which can be derived from physical observations
inside a dynamical family is the one of (423) Diotima, which has been
proposed in \citet{masiero2013} and \citet{paperIV}, and appears as a subfamily of
the family of (221) Eos formed by C-type interlopers\footnote{The
  membership of this ``Diotima'' family cannot be sharply defined, but
  if the dark interlopers of 221 with $3.05<a<3.10$ au were all
  members of family 423, then $f_v=0.1170$.}.

The family of (686) Gersuind is characterized by an intermediate
albedo which does not allow a reliable subdivision into bright and
dark. Apart from these two exceptions, for which an explanation is
available, all the cratering families of Table~\ref{tab:frac} appear
to be homogeneous, in the sense that they contain an acceptable
fraction of interlopers.

Indeed, families are statistical entities defined by a contrast of
density, formed by spreading fragments in a large volume of proper
element space which could not have been empty before: these
preexisting asteroids remain as interlopers, and can be detected by
physical observations when the background is of a different taxonomic
type from the prevalent family one. Taking into account that the
membership of our ``dynamical'' families has been determined using
only their proper elements, the a posteriori verification that the
families are compositionally homogeneous, as much as can be expected,
is a validation of our procedure; it confirms not only the existence
of the families we have proposed, but also that our methods to
terminate the aggregation of families are effective for our declared
purpose of providing a reliable membership. We note that it would be
relatively straightforward to use some method of classification giving
many more members to the families, but then this property of
compositional homogeneity could get lost.

\section{Properties of cratering families}
\label{s:properties}

As discussed in Section~\ref{s:intro}, one of our goals is to show
that not only is the composition homogeneous, but also the shape (in
the proper element space) of the families is consistent with a
possible original (immediately after the collision) distribution of
the escape velocities of the fragments.

Please note that we cannot use the information from the distribution
for the proper $a$ of the family members because these are affected by
nongravitational Yarkovsky \citep{paperIII} and YORP
\citep{paolicchiYORP} effects acting over the age of the
family. Therefore the distribution of proper $a$ does not at all
represent the original distribution of relative velocities in the
direction along track (apart for the case of recent families, with age
$<10$ Myr), but mostly contains information which can be used to compute
the ages.

On the contrary, the proper elements $e$ and $\sin{I}$ computed today
may preserve information on the original distribution of escape
velocities, mostly in directions orthogonal to the velocity of the
parent body. This should occur in most cases, although not for all
families, because there are dynamical mechanisms allowing a secular
change also of proper $e$ and/or $\sin{I}$. Moreover, it is necessary
to remember at this point the distinction between dynamical and
collisional families: for example, if a dynamical family contains two
collisional families, there have been two collisions, therefore two escape
velocity distributions contributing to the present shape of the
family.

\subsection{Asymmetry of velocity distribution}
\label{s:anisotropy}

Table~\ref{tab:anisotr} contains information, for each cratering
family we have identified, on the distribution of the differences
$\delta e$ and $\delta \sin{I}$ in the proper elements $e, \sin{I}$
between the fragments and the current position of the major body. More
precisely, the columns list family number, mean differences in $e,
\sin{I}$, corresponding STandard Deviations, Converted Escape
Velocity, and notes (explained in Section~\ref{s:list}). Given that
the families are all of cratering type, that is, the fraction of
fragments is $<1/4$ in volume (even less in mass, since the fragments
should be more fractured, and therefore less dense) the difference between the
current position (in proper $e, \sin{I}$) of the family namesake and
the original position of the parent body should be significantly
smaller.

For each family we have computed the two lowest-order moments of the
$\delta e$ and $\delta\sin{I}$ distribution, namely mean and variance,
and we show in the table the mean and the standard deviation for each
of the two proper element distributions, centered at the parent body.
Higher moments of these distributions could be computed easily, but
let us first understand the information contained in the first two.

\begin{table}[h!]
\centering
\footnotesize
\caption{Asymmetry of cratering families.}
\label{tab:anisotr}
\medskip
\begin{tabular}{|l|r|r|r|r|r|l|}
\hline
NC &  $Mean(\delta e)$ &  $Mean(\delta sI)$ & $STD(\delta e)$ & $STD(\delta sI)$& CEV & Note\\
\hline 
3   &  0.0013  &  0.0019  &  0.0028  &  0.0023& 0.0055& 2 ag?\\
3+  &  0.0011  &  0.0014  &  0.0026  &  0.0019& 0.0055& \\
3-  &  0.0015  &  0.0023  &  0.0030  &  0.0026& 0.0055&\\
4   &  0.0000  &  0.0042  &  0.0091  &  0.0054& 0.0122&\textbf{2 ag}\\
4+  &  0.0044  &  0.0012  &  0.0099  &  0.0046& 0.0122&\\
4-  & -0.0034  &  0.0066  &  0.0066  &  0.0048& 0.0122&\\
5   & -0.0037  & -0.0030  &  0.0131  &  0.0161& 0.0025& \\ 
10  & -0.0056  &  0.0013  &  0.0134  &  0.0048& 0.0104&\\
15  &  0.0021  &  0.0016  &  0.0115  &  0.0083& 0.0052&\textbf{2 ag}\\
15+ &  0.0037  & -0.0017  &  0.0109  &  0.0074& 0.0052& \\
15- &  0.0012  &  0.0035  &  0.0118  &  0.0082& 0.0052&\\
20  &  \textbf{0.0004}&\textbf{0.0008}&  0.0050  &  0.0021& 0.0029&2 ag?\\
31  & -0.0144  &  0.0015  &  0.0204  &  0.0034& 0.0058& HI\\
31+ & -0.0002  &  0.0010  &  0.0125  &  0.0041& 0.0058& \\
31- &\textbf{-0.0305}&0.0021& 0.0148&0.0023& 0.0058& \\
87  &  0.0074  &  0.0004  &  0.0063  &  0.0022& 0.0066&OS\\
163 & -0.0019  &  0.0046  &  0.0035  &  0.0035& 0.0015& \\
283 &  0.0018  &  0.0001  &  0.0040  &  0.0022& 0.0029&\textbf{2 ag}\\
283+&  0.0010  & -0.0007  &  0.0032  &  0.0013& 0.0029&\\
283-&  0.0037  &  0.0018  &  0.0047  &  0.0027& 0.0029&\\
569 & \textbf{-0.0001}& \textbf{-0.0002}&0.0024&0.0018& 0.0015&\textbf{sub}\\
686 &  0.0021  & -0.0012  &  0.0081  &  0.0044& 0.0012&\\
\hline
302 &  0.0007  &  0.0002  &  0.0017  &  0.0008& 0.0007&  Y\\
396 & -0.0002  & -0.0011  &  0.0011  &  0.0008& 0.0009&  Y\\
606 &  0.0018  & -0.0003  &  0.0007  &  0.0005& 0.0008&  Y\\
\hline
363 &  0.0050  &  0.0034  &  0.0078  &  0.0042 & 0.0020&\\
480 &\textbf{0.0299}& -0.0017&\textbf{0.0237}&0.0030& 0.0013& HI\\ 
480+&\textbf{0.0204}& -0.0026&\textbf{0.0188}&0.0025& 0.0013&\\
480-&\textbf{0.0384}& -0.0009&\textbf{0.0245}&0.0032& 0.0013&\\
1303&  0.0014  &  0.0009  &  0.0083  &  0.0056 & 0.0020& HI\\
1547& -0.0001  &  0.0002  &  0.0007  &  0.0004 & 0.0005& R\\
\hline
96  & \textbf{-0.0004}& \textbf{-0.0004}&0.0022&0.0020& 0.0038&\\
148 & -0.0001  & -0.0008  &  0.0071  &  0.0017& 0.0022& HI\\
179 & \textbf{-0.0076}  & -0.0039  &  0.0061  &  0.0026&0.0017& \\
179+& \textbf{-0.0083}  & -0.0041  &  0.0059  &  0.0025&0.0017& \\
179-& -0.0039  & -0.0022  &  0.0057  &  0.0027&0.0017& \\
410 & -0.0032  & -0.0075&0.0051&0.0030&0.0024&\\
778 & -0.0061  & -0.0024  &  0.0040  &  0.0018 &0.0026& R\\
1222&  0.0042  & -0.0004  &  0.0117  &  0.0026 &0.0007&2 ag?,HI\\
1222+&-0.0015  & -0.0023  &  0.0081  &  0.0016 &0.0007& \\
1222-& \textbf{0.0087}&0.0012&0.0122&0.0022&0.0007& \\
\hline
\end{tabular}
\end{table}

The fact that these distributions of $\delta e$ and $\delta\sin{I}$
are representative of the original ejection velocities of the
fragments is not an assumption but something to be tested case by
case. In particular, a necessary condition for this interpretation to
be legitimate is that both the mean and the standard deviation are of
the same order of the escape velocity from the parent body. Therefore,
in the table we also list the estimated escape velocity, converted
into $\sqrt{(\delta e)^2+ (\delta\sin{I})^2}$ by the \citet{zappala90}
metric (converted escape velocity, CEV). Please note that this
conversion is valid only ``on average'', because the actual value of
the difference in orbital elements due to a given escape velocity
(outside of the gravitational sphere of influence of the parent body)
actually depends on short-periodic quantities such as the true anomaly
$f$ and the argument of latitude $\omega +f$ at the time of the
impact. For this reason, and also because the masses of the parent
bodies are only approximately known, a difference between the mean
$\delta e$, $\delta \sin{I}$ and the CEV, by a factor of approximately
$2$, is not a problem, while differences of an order of magnitude
require an explanation.

In principle, values of the mean and/or of the STD of $\delta e$
and/or $\delta\sin{I}$ which are larger (by a significant factor) than
the conversion of the escape velocity, imply some dynamical
evolution occurring after the family formation. The dynamical evolution
mechanisms, acting on proper $e$ and/or $\sin{I}$ of a large fraction
of family members, would then need to be identified; otherwise, the
family definition should be reconsidered.

On the other hand, values that are too small, such as those one order
of magnitude smaller than the conversion of the escape velocity, are
unlikely to occur by chance in a single collisional process. The
reason for this is simply the formula
$V_\infty=\sqrt{V_0^2-V^2_{esc}}$, where $V_0$ is the launch velocity
of the fragment from the crater, $V_{esc}$ is the escape velocity from
the surface of the parent body, and $V_\infty$ is the velocity of the
fragment after leaving the gravitational well of the parent body. This
means that values of $V_\infty$ much lower than $V_{esc}$ are
unlikely, although they are not strictly impossible\footnote{As an
  example, there could originally have been an asymmetry mostly in
  $\delta a$, which is not now measurable because of non-gravitational
  effects.}. Therefore these cases should be investigated for possible
interpretation as families generated by two (or more) cratering
events.

Indeed, from the computation of the ages of the families (see
Appendix~\ref{s:age}), we do know some cases of families with two
different ages, computed from the IN (smaller proper $a$) and OUT
(larger $a$) side of the V-shape in the $(a, 1/D)$ plane. This
includes three cratering families: 4, 15, and 283; moreover, we know
two dubious cases for the cratering families 3 and 1222. In these cases
we have therefore computed separate values of mean and STD for the proper
$a>a_0$ and for the $a<a_0$ portions of the family, where $a_0$ is the
proper $a$ of the parent body.

We have not included the family of (2) Pallas in this table, because
it is too depleted by mean motion resonances (of the three-body type),
which at such a high inclination ($\sin{I}\sim 0.54$) open large
gaps. 

We now briefly commenting on some cases with asymmetry that is either
too large or too small.


\subsection{Families with two ages}

For the three cratering families (4, 15, 283) already known to have
two ages (from the different slopes of the V-shape in $a$) the
separation of the IN and OUT sides of the family (see the symbols such
as $4+$ and $4-$ in the first column of Table~\ref{tab:anisotr})
nicely confirms, with clearly distinct mean values, the existence of
two separate jets with very different directions in proper element
space: see for example, Figure~\ref{fig:283_aI}, which visually confirms the
interpretation of the dynamical family of (283) Emma as two well
separated jets. The same applies to the family of (4) Vesta; see
\citet{bigfamilies}[Figure 12].

\begin{figure}[h]
\centering \figfigincl{12 cm}{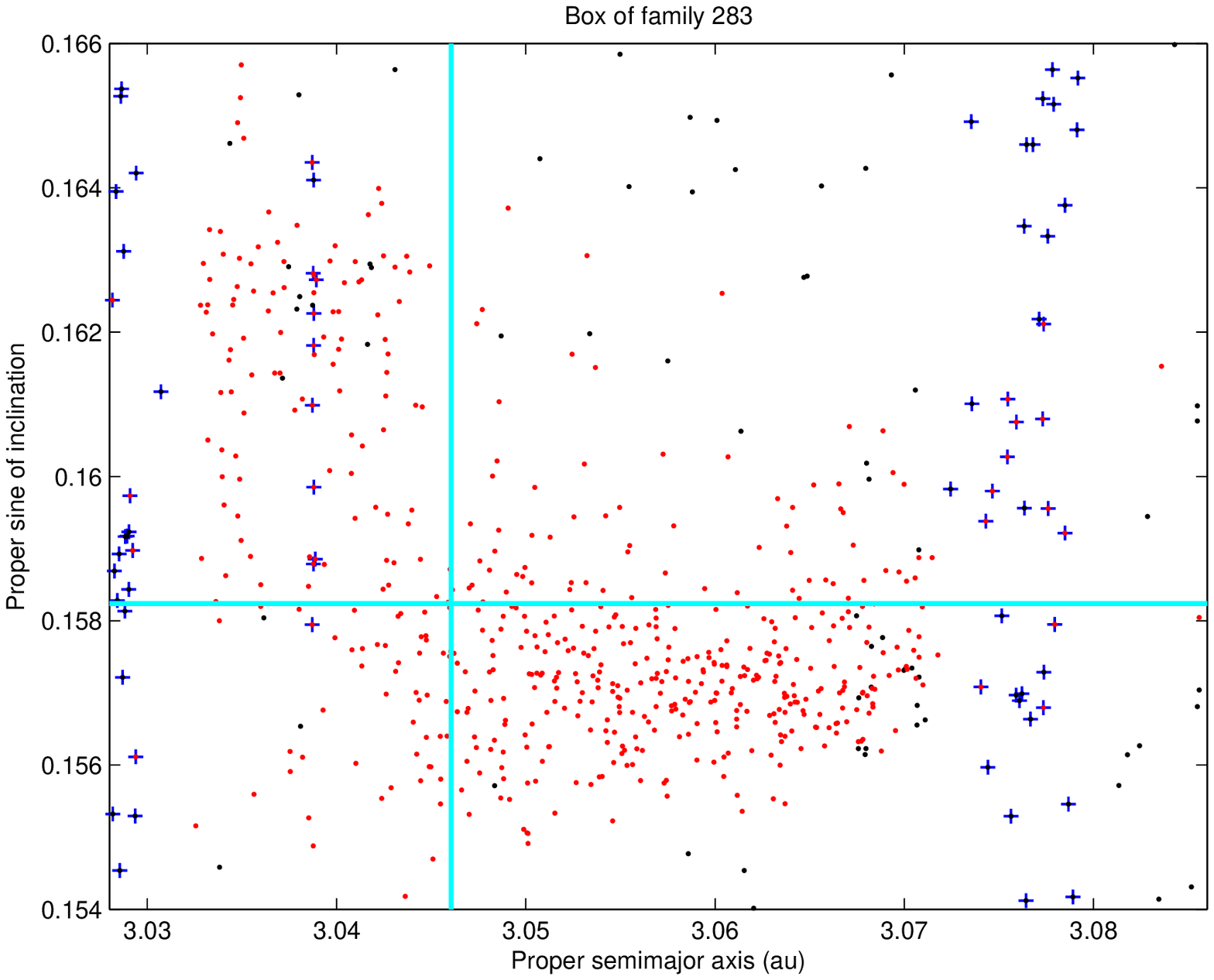}{Family of (283) Emma projected
  on the proper $(a, \sin{I})$ plane. Red points represent family
  members, black points are background objects and blue crosses are
  objects in mean motion resonances. Position of (283) is marked by a
  large cyan cross. The two separate jets indicate a double collision
  origin.}
\end{figure}

For the family of (3) Juno, which was a dubious case of two ages from
the V-shape \citep{paperIII}, the data on the asymmetry do not confirm
the interpretation as two separate collisional families. The values
for the momenta of the distribution of proper elements are somewhat
low, although by less than a factor $3$ for $\sin{I}$, and there is no
improvement if the family is split into IN and OUT side (see rows marked
$3+$ and $3-$). Given that the significance of the separation into two
ages is marginal (about $1.2$ STD; see Table~\ref{tab:slope_crat}),
this case remains dubious.

For the family of (569) Misa, as shown in Table~\ref{tab:anisotr}, the
asymmetry is indeed too low in $e,\sin{I}$ for it to be
the outcome of a single cratering event. On the other hand, we do know that
this family was formed by two separate collisions; see the figure in
the $(a, 1/D)$ plane showing a W shape rather than a V
\citep{paperIV}[Figure 1]. There is therefore another collisional family,
much younger than the other one (see Table~\ref{tab:age_crat}). We
have separated the family 15124 from 569 using this W-shape; see the top panel of
Figure~\ref{fig:15124_Ie}.  Admittedly, a few
members from 569 could be attributed to 15124, but this does not affect the
results because our 15124 family has many more members.

\begin{figure}[h!]
\centering
\figfiginclnocap{11.5 cm}{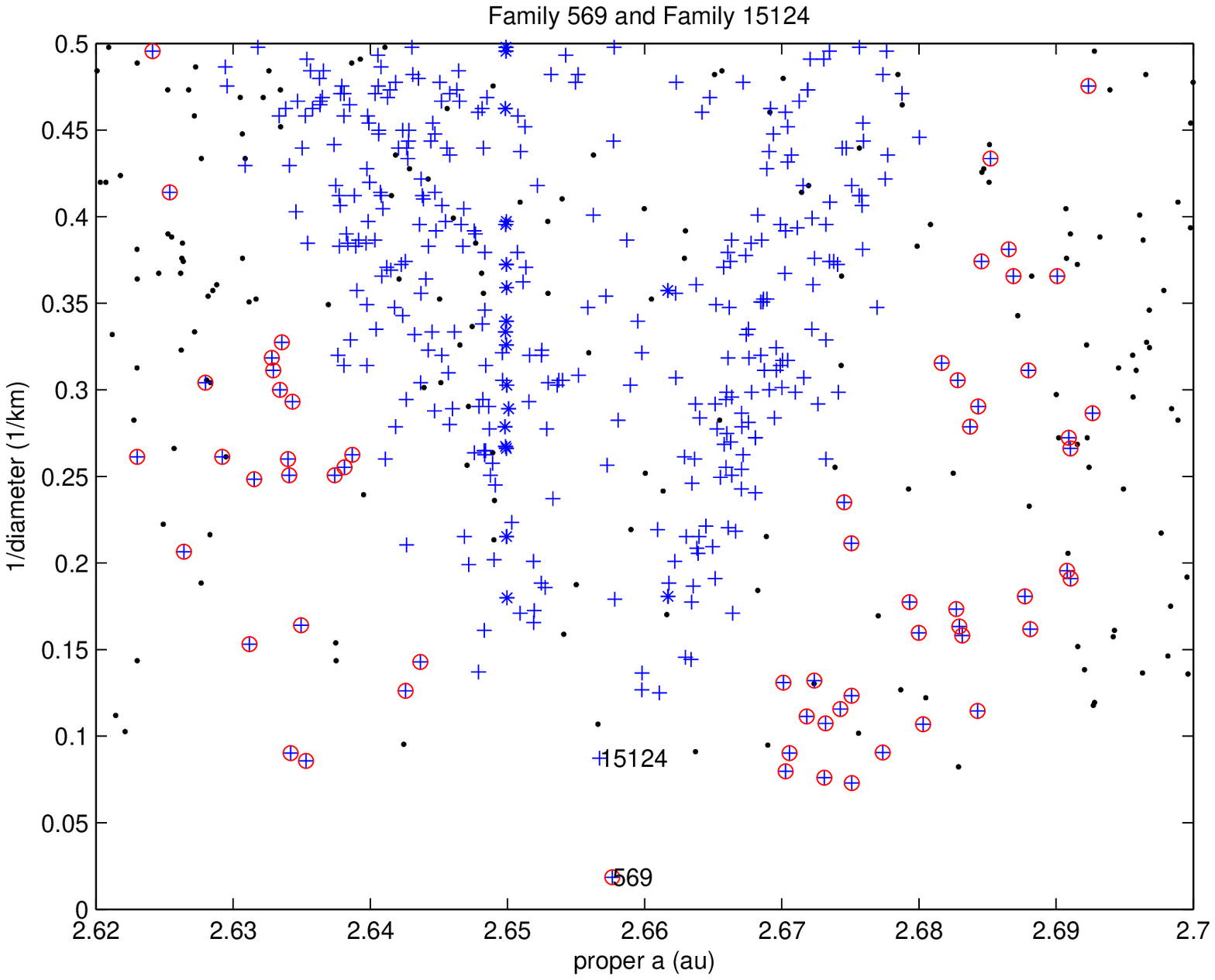}

\figfigincl{11.5 cm}{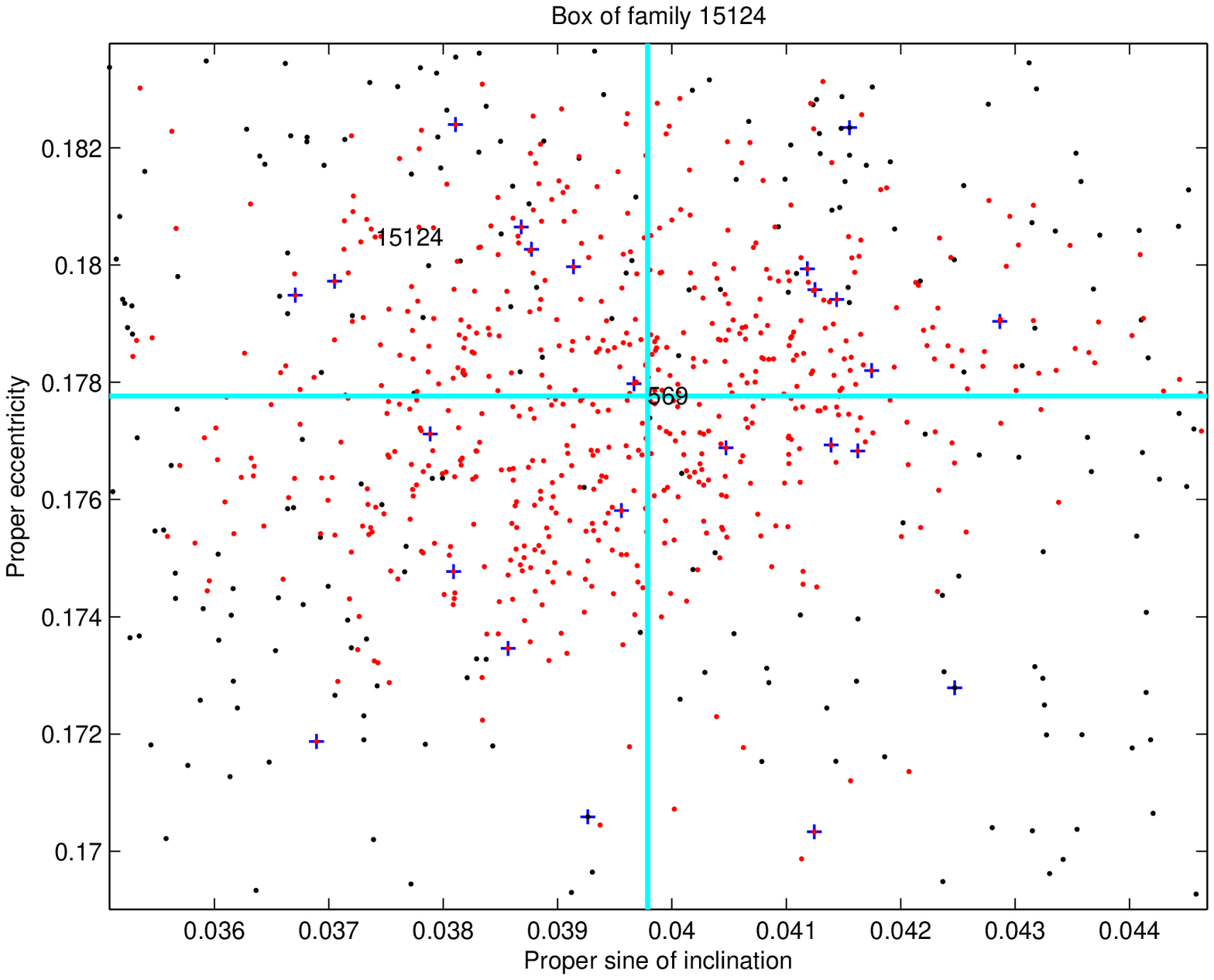}{Top: family of (569) Misa is
  separated into a family 569 with $70$ members (crosses encircled in
  red) and a family 15124 with $577$ members (blue crosses; stars are
  also in mean motion resonances). Bottom: distribution of the family
  15124 in the proper $(\sin{I},e)$ plane, showing the central
  position of (569) (cyan cross) and the peripheral location of
  (15124), not necessarily representative of the position of the
  parent body.}
\end{figure}

If family 15124 were the outcome of a second cratering on (569), then
it should have shown an appropriate asymmetry of $\delta e$ and
$\delta\sin{I}$ with respect to (569). As a matter of fact, as shown
in the bottom panel of Figure~\ref{fig:15124_Ie}, (569) is right in
the center of the distribution of 15124 in the proper $(\sin{I},e)$
plane. For the family 15124 the mean differences with respect to (569)
are $mean(\delta e)=-0.00004$, $mean(\delta\sin{I})=-0.003$, by far
too small with respect to the escape velocity from (569), with
CEV$=0.0015$. From this we conclude that family 15124 is not a second
cratering but a fragmentation. The total volume of the known members
of 15124 corresponds to a sphere of diameter $D\simeq 25$ km, too
large to be an ejecta from a crater on (569), which is a body with
$D\simeq 73$ km.

Neither the second cratering on (569) nor the fragmentation of a first
generation fragment from (569) are therefore possible. The only other
explanation which we could find is that family 15124 is a complete
fragmentation of a background object, which by unlikely coincidence
has a center in proper element space very close to the location of
(569). Following a well known argument\footnote{"How often have I said
  to you that when you have eliminated the impossible, whatever
  remains, however improbable, must be the truth?" Sherlock Holmes to
  dr. Watson, in: C. Doyle,\emph{The Sign of the Four}, 1890, page
  111.} we are obliged to accept the latter explanation.

The 15124 family therefore appears to have nothing to do with (569),
not even as a second-generation parent. The albedo of (569) is
$0.0297\pm 0.001$ \citep{IRAS}, and that of (15124) is $0.077\pm 0.007$
\citep{WISE}, both well within the C-type range, but different enough
to support the model we are proposing of a fully independent
fragmentation family which is by chance overlapping a cratering
family.

In the dubious case of the family of (1222) Tina, discussed in
\citet{paperV}, the hypothesis of a double collision origin is
confirmed, as shown in the rows marked $1222+$ and $1222-$, by the
presence of two very distinct jets. The value of $mean(\delta e)$ is
very high, but this can be interpreted as being due to the $g-g_6$
secular resonance, which affects the family \citep{carrubamorby}, in
particular in the form of anti-aligned libration states. This family
still only has $107$ members, and therefore no statistically robust
conclusions can be made at this stage.


After handling these double-collision cases, the most prominent cases
of anomalous values, both larger and smaller than the CEV by factors
more that four, are shown in bold in Table~\ref{tab:anisotr}.

\subsection{Overly large asymmetries}

The most striking case of overly large value is the mean of $\delta
e$ for the family of (31) Euphrosyne. Since the asymmetry of the
proper element distribution in the IN and OUT portions is obvious, in
particular in the proper $e$, as shown by
Figure~\ref{fig:31_ae_meane}, we have computed the separate values for
the two sides; see the rows $31+$ and $31-$. However, this separation
has made the situation worse, in that the $mean(\delta e)$ limited to
the members with $a<a_0$ has grown to the value of $-0.03$, which is
more than five times the value corresponding to the estimated escape
velocity. In Figure~\ref{fig:31_ae_meane} we have plotted a running
mean of the proper $e$: it shows that the distribution of $e$ for
$a<a_0$ is completely different from the one for $a>a_0$. This is even
more interesting considering that the V-shape in $(a, 1/D)$ does not
indicate two different ages for this family; see
Table~\ref{tab:slope_crat} and the top panel of
Figure~\ref{fig:31_vshapea_outliers}. This case needs to be
investigated to find a suitable explanation, acting mostly on proper
$e$, the asymmetry in $\sin{I}$ being less than the CEV (see
Section~\ref{s:shapes}).

\begin{figure}[t!]
\centering
\figfigincl{12cm}{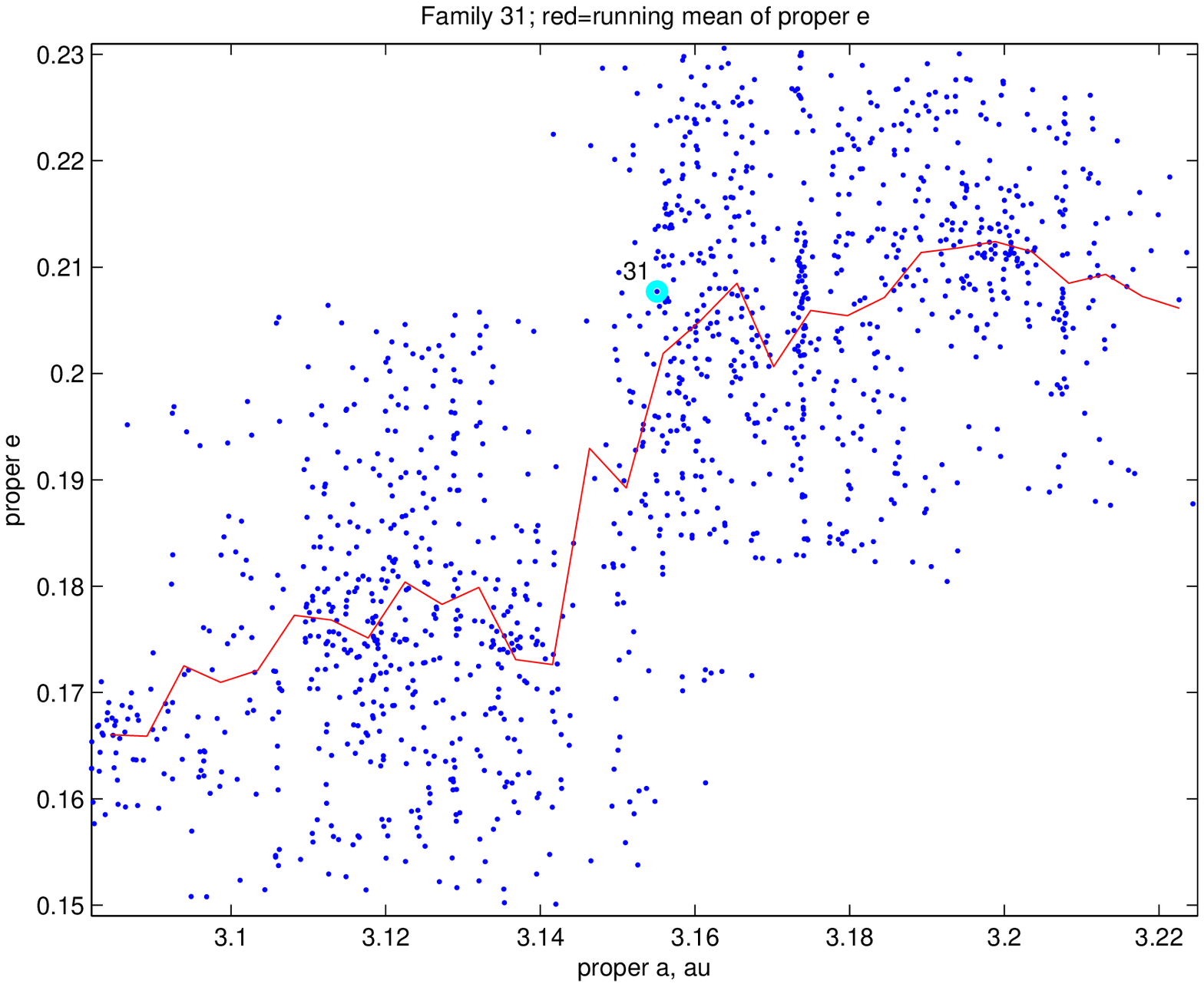}{Family of (31) Euphrosyne projected on
  the proper $(a, e)$ plane; the big blue dot shows the position of
  the parent body (31) Euphrosyne. The red line is the running mean of
  the values of proper $e$ vs. proper $a$, for the members of the
  family, showing the substantial jump to lower values in the IN side
  of the family.}
\end{figure}
\nobreak The second largest asymmetry is in the $\delta e$ of the
family (480) Hansa: we had thought this depended upon lower accuracy
of the proper $e$ computed in this region, with large proper $\sin{I}$
and low proper $e$. This lower accuracy resulted from a failure to
correctly identify the proper mode because other forced oscillations
are larger \citep{CarMish}. After improving proper $e$ using a form of
frequency analysis, that is using as proper $e$ the amplitude of the
highest peak in the spectrum of $e\sin(\varpi)$ in the range of
periods between $50,000$ and $70,000$ y, the asymmetry was slightly
increased. Therefore, the quality of the proper elements has been
improved, especially for very low proper $e$, but this was not the
problem.
Neither mean nor RMS of $e$ can be explained by ejection velocity,
being larger by more than an order of magnitude, or by mean motion
resonances in the model (there is $11/4$J at $a=2.649$ au, but its
effect is limited to a small portion of the family; see
Figure~\ref{fig:480_ae_synt} (top)).  We tried the separation IN/OUT,
and found that the asymmetry is much more pronounced in the portion
with $a<a_0$, with $mean(\delta e)$ as large as $\sim 30$ times the
CEV. The value of $mean(\delta e)$ is also too large also
$a>a_0$. There is less asymmetry in $\delta \sin{I}$, but also in this
the two sides are different. This case therefore also needs an
explanation for the peculiar $\delta e$ distribution (see
Section~\ref{s:shapes}).

In the family of (179) Klytaemnestra, to explain the large asymmetry
in proper $e$, we tried the separation IN/OUT, but the situation did
not improve, actually the $mean(\delta e)$ value in the row $179+$ is
even larger than the one for the entire family, and is approximately
five times the CEV. Moreover, as already
mentioned in \citet{paperV}[Sec. 5.3], the V-shape in $(a,1/D)$
appears impossible to model collisionally.  The interaction with the
Eos family has to be considered to provide an explanation, see
Section~\ref{s:shapes}.

There are two cases in which the asymmetry is marginally high, for
example between $3$ and $4$ times the CEV, namely families 410 and
163.

One unusual feature of the family of (410) Chloris is that the proper
$\sin{I}$ of the parent body (410) is the top value in the family,
which contributes to the comparatively large negative
$mean(\delta\sin{I})$. We note that there is another family, (32418)
2000 RD33, separated from 410 in proper $a$ by the three-body
resonance 3J-1S-1A.  The range of values of both proper $e$ and
$\sin{I}$ for 32418 is included in the corresponding ranges for
410. Therefore we have considered the possibility that these two
families arise from a single cratering event on (410): the merged
family would still be of cratering type, and would have a somewhat
smaller asymmetry in $\delta\sin{I}$. However, we do not think there
is convincing evidence for this merger; for example, the average WISE
albedos of the two families are both uncertain and only marginally
consistent\footnote{We have tried computing an age for the merged
  family, which would have $201$ members. There in only one slope, the
  OUT one, but the fit does not appear good enough.}. We can only wait
for more data to make a decision, either as more and more accurate
physical observations or as new family members. The family (410)
without this merge has only $120$ members, and therefore we do not
think it is useful to investigate this case in depth.

In \citet{paperIII} the two families of (163) Erigone and (5026)
Martes have been considered together for the purpose of computing an
age, since they form a single V-shape with the same age; see
Table~\ref{tab:age_crat}. We have therefore merged them for computing
the entry in Table~2: together they give a somewhat large asymmetry,
with $mean(\sin{I})$ equal to $3.07$ times the CEV. If the family of
(163) were considered separate from the one of (5026), the largest
asymmetry parameter, which is $mean(\delta\sin{I})$, would increase
from $3.07$ to $3.33$ times the CEV. Therefore the analysis of the family
shape in proper $(e,\sin{I})$ supports the merger. This case should be
analyzed to provide an explanation for the peculiar $\delta e$
distribution (see Section~\ref{s:shapes}).

\subsection{Overly small asymmetries}

One possible solution for the cases in which the asymmetry parameters
are too low is to find indications for a possible double
collision. There is one case in which this is possible: for the family
of (20) Massalia. Indeed, the asymmetry is somewhat low, equal about
one third of the CEV for $\delta \sin{I}$.  However, a close look at
the distribution of the proper elements of the family members,
especially in the proper $(\sin{I},e)$ plane (Fig.~\ref{fig:20_Ie}),
shows a halo at lower $e$, higher $\sin{I}$. This indicates a second
collisional family, with positive $mean(\delta\sin{I})$ and negative
$mean(\delta e)$, compensated for the larger collisional family with
opposite asymmetry.  Unfortunately, the second family appears to be
heavily superimposed, to the point that we currently can neither
separate it nor compute a second age.

\begin{figure}[t!]
\centering \figfigincl{12 cm}{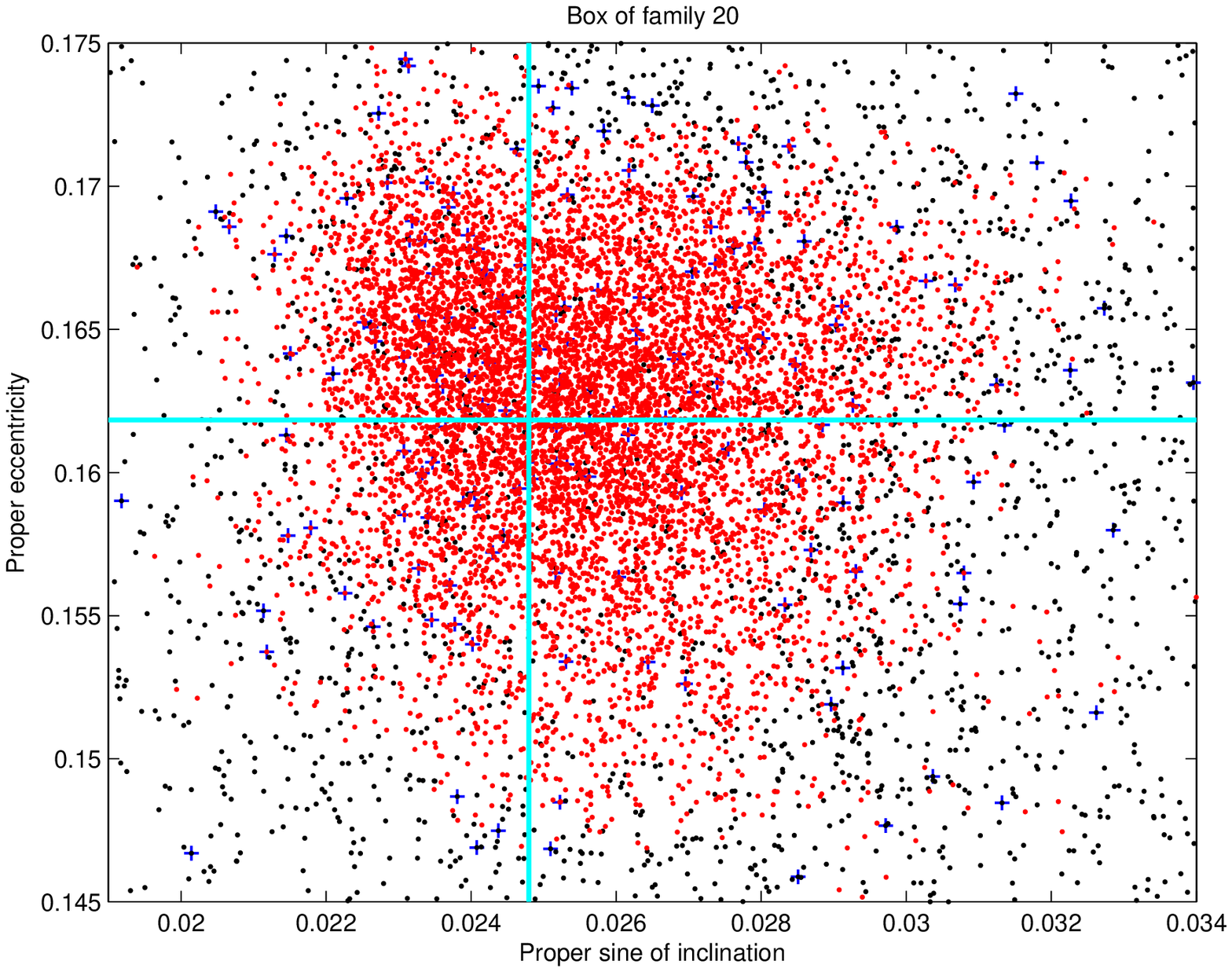}{Family of (20) Massalia projected
  on the proper $(e, \sin{I})$ plane, showing a halo in the lower
  right quadrant (with respect to the position of the parent body,
  marked by the cyan cross). Points are marked as in
  Fig.~\ref{fig:283_aI}.}
\end{figure}

For the only other case with overly low asymmetry (by a factor $~9$ in
both coordinates), the family of (96) Aegle, we currently have no
explanation. On the other hand, this family has too few members (only
120) to understand details of its structure. A solution of this case
may be found when the number of members has  grown to $\sim 200$
or more.

\section{Family shapes in need of dynamical explanation}
\label{s:shapes}

After solving, or at least proposing a solution for most of the cases
of anomalous asymmetry, we are left with only three cases with overly
large asymmetry, namely families 31, 480, 179, and one case with
marginally large asymmetry, family 163. In this section, we discuss
possible collisional and/or dynamical interpretation of these four
cases.

\subsection{Family of (31) Euphrosyne}

The main dynamical feature of family 31 is the presence of the
strongest linear secular resonance $g-g_6$ which cuts the family; see
\citet{carrubaVN2018}[Figure 4].  According to our new computations of
the location of the secular resonances, family 31 is cut by the
$g-g_6$ into two parts, very nearly corresponding to the IN and OUT
sides with respect to the parent body.  Figure~\ref{fig:31_aI_syntg6}
shows the resonance strip, which has been computed by means of a new
synthetic theory for the secular frequencies $g$ and $s$; the fixed
value of the third coordinate is $\sin{I}=0.45$ for the top plot and
$e=0.19$ for the bottom one. 

The method to compute these frequencies
as a smooth function of proper $(a,e,\sin{I})$ is a generalization of
the one used in \citet{trojans2}, and is fully described in Kne\v
zevi\'c and Milani (2018, \textit{in preparation}).
The modeling of $g,s$ as smooth functions (in fact polynomials) is
possible only after removing the asteroids in the region that are
strongly affected by mean motion resonances, as detected by the
estimated Lyapounov Characteristic Exponent and/or instability of
proper $a$. The lines drawn in the figure are level lines of the
best fit polynomial representation of $g-g_6$.

\begin{figure}[h]
\centering
\figfiginclnocap{11.5 cm}{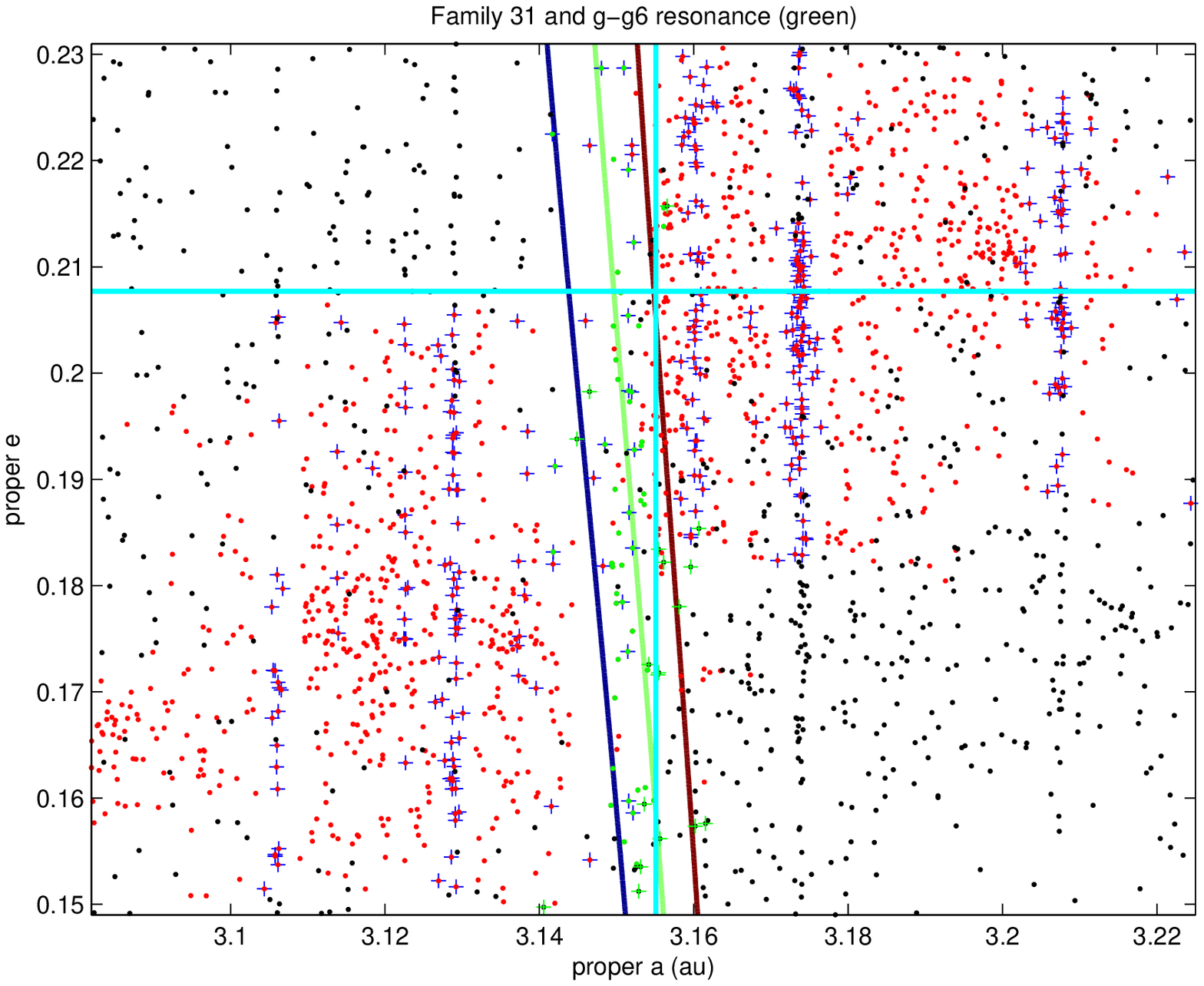}

\figfigincl{11.5 cm}{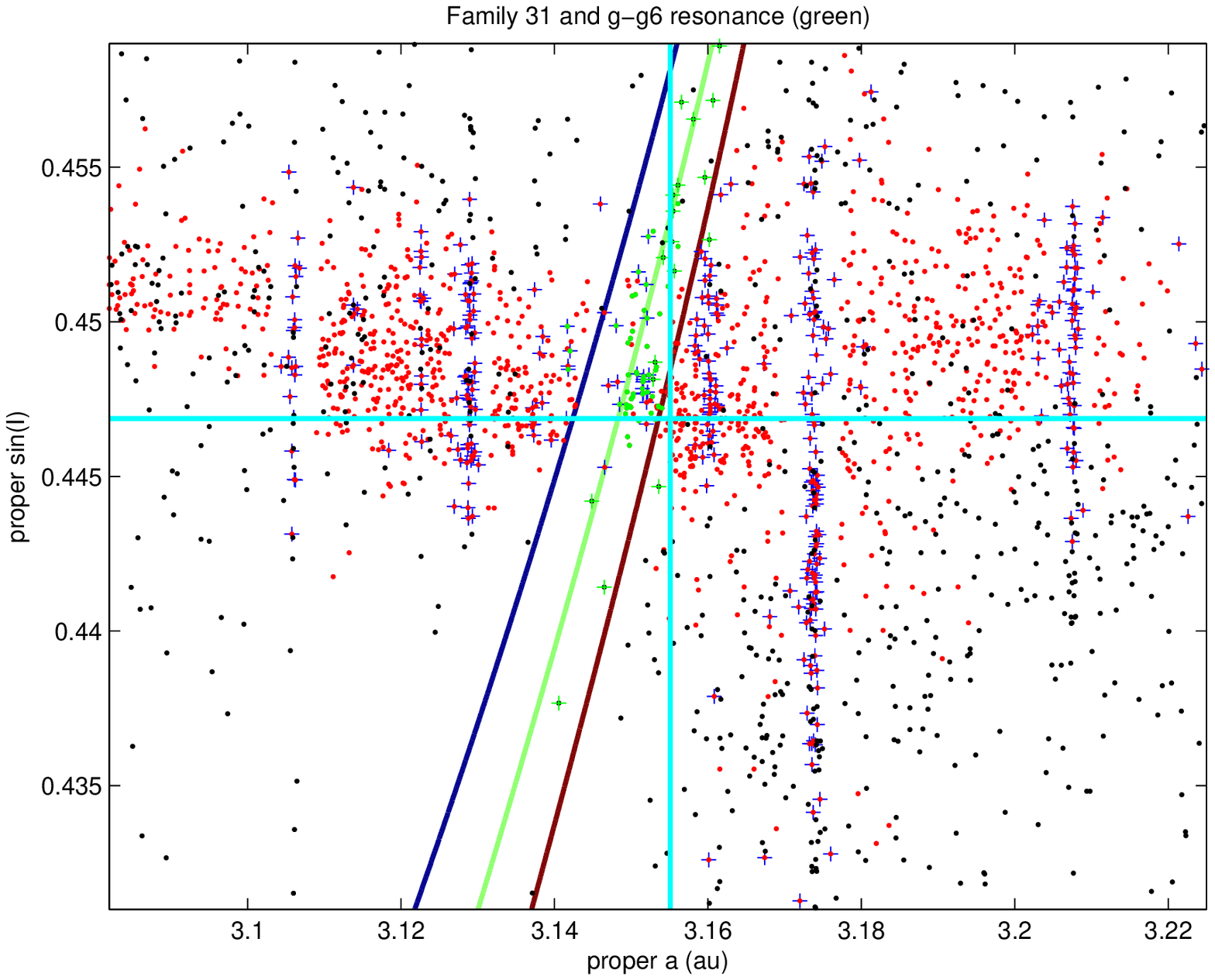}{Family of (31) Euphrosyne projected
  on the proper $(a, e)$ plane (top) and the proper $(a,\sin{I})$
  plane (bottom).  The $g-g_6$ resonance is described in two ways,
  with green points for individual asteroids with their synthetic
  proper frequencies $g$ such that $|g-g_6|<2$ arcsec/y, and with
  contour lines of the synthetic theory of proper frequencies with
  values $(-2, 0, +2)$ arcsec/y. Red points are family members, black
  are background objects found in the family box, and blue crosses
  indicate chaotic orbits. The parent body (31) is marked by the cyan
  crosses.}
\end{figure}

The parent body (31) Euphrosyne is very close to the secular
resonance, therefore fragments ejected from it with $\delta a=a-a_0<0$
in most cases end up in the $g-g_6$. Even if the ejection velocity is
large enough to put the fragment on the other side of the resonance,
the fragment is likely to have a secular $da/dt>0$ due to the
Yarkovsky effect. This is because the presence of a negative V-base
(see Figure~\ref{fig:31_vshapea_outliers}, top) in the V-shape plot
may indicate that $\delta a<0$ corresponds mostly to prograde spin
(see Appendix~\ref{s:paol}).
Therefore, even the fragments which are not originally inserted in the
secular resonance end up falling into it as a result of the Yarkovsky
effect, and in both cases most of them may have been pushed by $g-g_6$
to high eccentricities (such as $0.5$), leading to close approaches to
Jupiter and subsequently to ejection from our solar system, and the
formation of interstellar asteroids. The occurrence of instability for
ejecta from (31) has been studied with extensive numerical simulations
by other authors, such as \citet{masiero2015}, with the result that
$~80\%$ of the test particles entering the $g-g_6$ resonance are
evacuated from the family region into either near-Earth of
Jupiter-crossing orbits. Given our analysis below, this estimate of
survival rate may even be optimistic.\footnote{The $g-g_6$ resonance
  contains an island of relative stability, due to anti-aligned
  libration, similarly to what was reported for the Tina family
  \citep{carrubamorby}. However, this island has a small volume in
  proper element space, and therefore such a relative stability
  affects only a few members, and does not change the overall picture
  of a dominant strong instability.}

\begin{figure}[h]
\centering
\figfiginclrotnocap{10 cm}{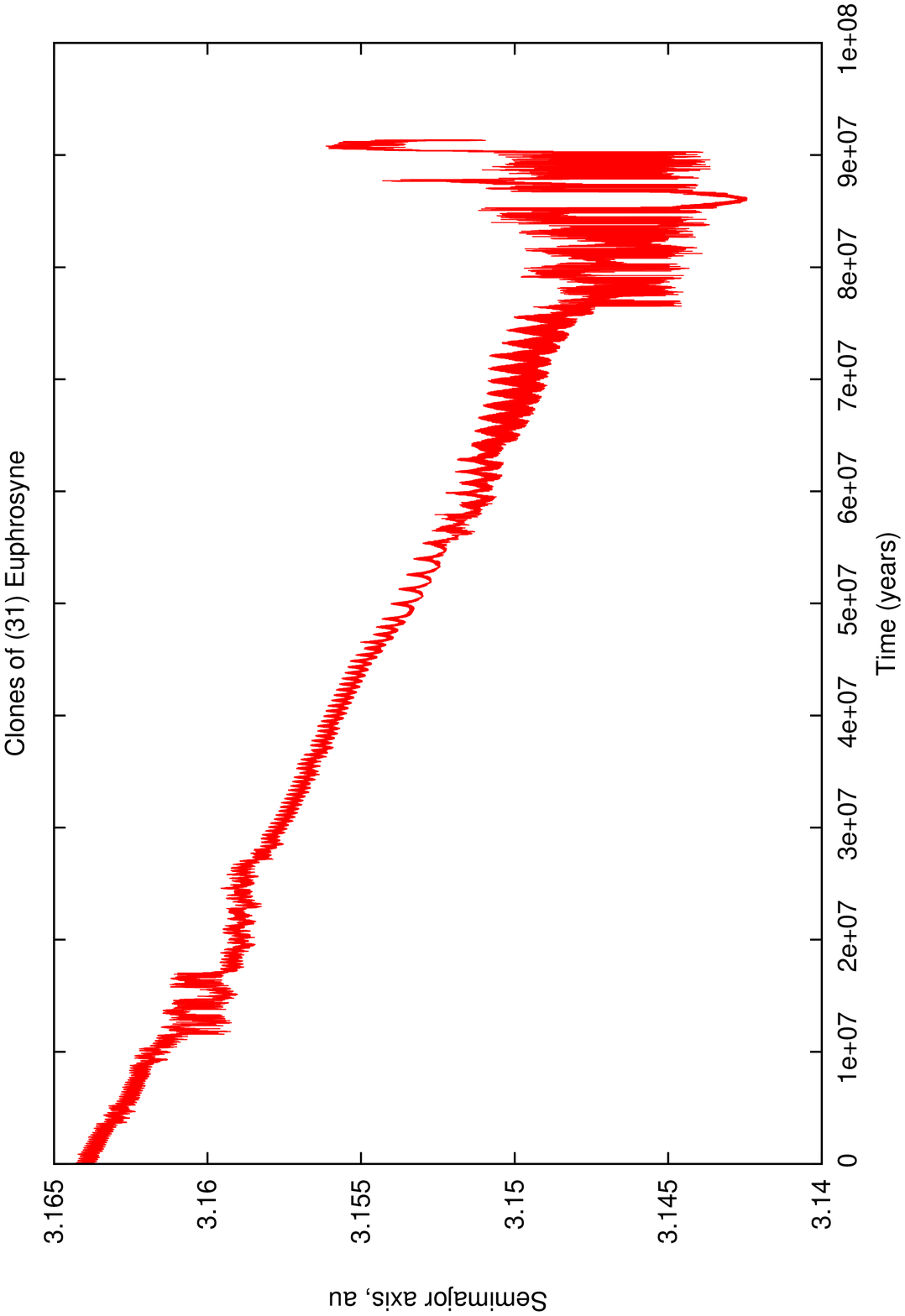}

\figfiginclrot{10 cm}{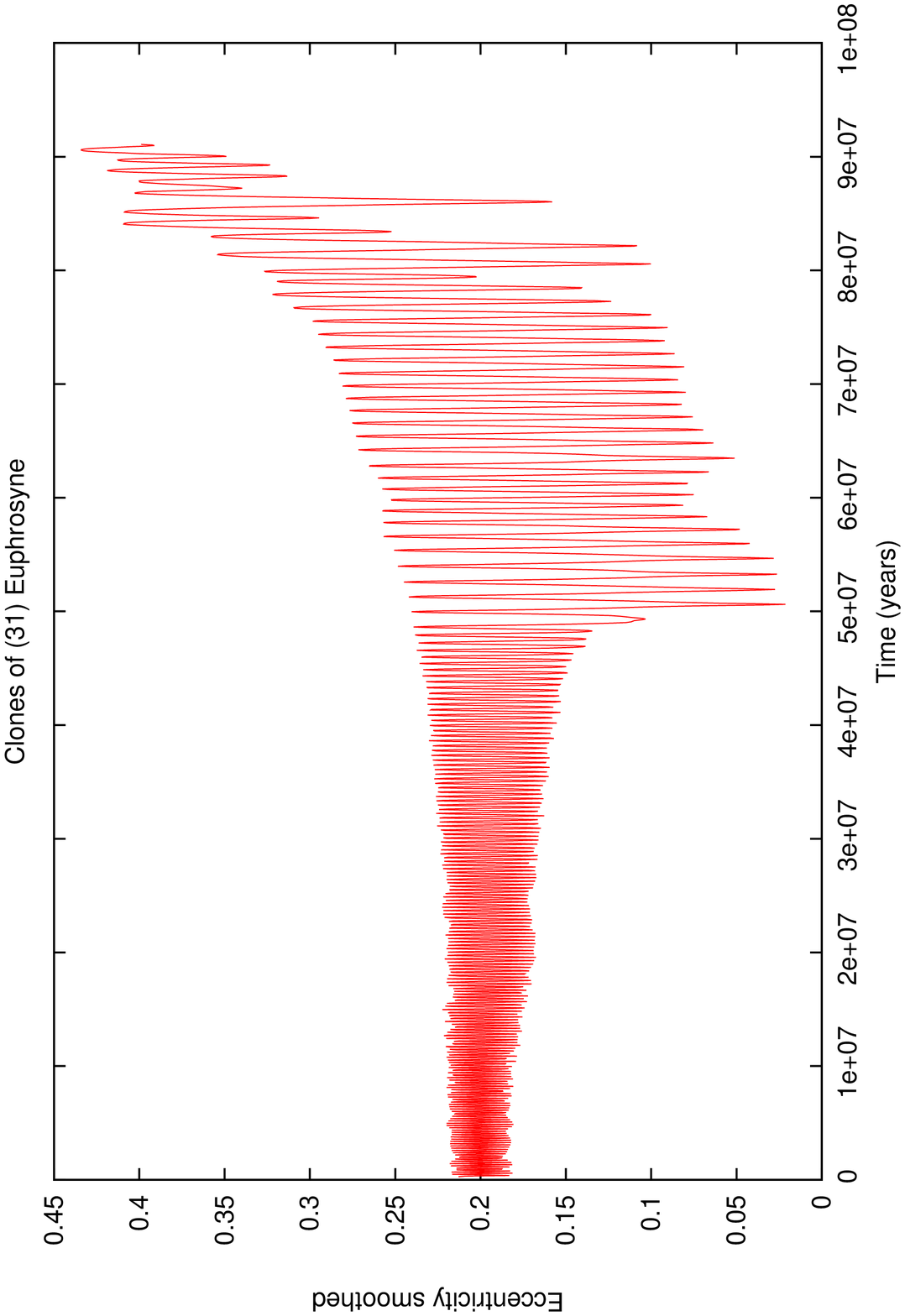}{Digitally filtered time series
  from the output of a numerical experiment for $100$ My, with initial
  conditions equal to those of (31) but for osculating $a=3.165$ au,
  and a Yarkovsky model such that the secular drift should be $da/dt=
  -2\times 10^{-4}$ au/Myr. Top: filtered semimajor axis (au). Bottom:
  filtered eccentricity, showing much larger oscillations after the
  entry into the secular resonance $g-g_6$ at about $+50$ Myr.}
\end{figure}

The fragments ejected with $\delta a>0$ are mostly with retrograde
spin and secular $da/dt<0$ due to Yarkovsky. Even in this case, there
are two possible outcomes: if they are ejected initially to a proper
$a> 3.174$ au, where the strong three-body resonance 5J-2S-2A is located,
they come back to lower $a$ until they meet this resonance, becoming
very strongly chaotic and also, in most cases, being ejected from our
solar system.  The fragments ejected with smaller $\delta a$ also
migrate to lower values of $a$ and therefore end up in the $g-g_6$
resonance. 

Figure~\ref{fig:600032_esmot} shows the output of a numerical
experiment which refers to this case: the initial osculating $a=3.165$
au is larger than that of the parent body ($a_0=3.1561$ au) and the
Yarkovsky secular drift is $da/dt=-2\times 10^{-4}$ au/Myr, a value
corresponding to a C-type asteroid with diameter $D\simeq 3$ km and
retrograde spin; see Table~\ref{tab:age_crat}.

 When the body enters the $g-g_6$ secular resonance (at $\simeq +50$
 Myr of the orbit propagation) the smoothed eccentricity at first
 oscillates between values very close to $0$ and $0.3$, and later
 between $0.1$ and $0.4$. At $+76$ Myr it also enters in the three-body
 resonance inside the secular one, and $e$ grows even more,
 oscillating between $0.2$ and $0.5$. Starting at $+84$ Myr there are
 close approaches to Jupiter, until the last at $+91.35$ Myr results in
 ejection to a hyperbolic orbit; this example ends up in an
 interstellar asteroid, after a time span an order of magnitude
 smaller than the age of family 31.

 As shown by this and other examples, entering into the $g-g_6$
 secular resonance, both from the lower $a$ edge and from the higher
 $a$ one, is like Swiss roulette\footnote{In the Russian roulette, as
   defined in 1840 by the Russian writer M. Lermontov, one of the
   chambers is loaded, the other five are empty: the game of pulling
   the trigger with the gun pointed to one's head is very
   dangerous. In the Swiss version, introduced in 1937 by the Swiss
   writer G. Surdez, five of the six chambers are loaded, and the game
   is very likely to be fatal.}, in which most of the objects are lost
 in interstellar space.  The dynamics due to this interaction of
 resonances and Yarkovsky effect is too complex to be quantitatively
 modeled, but from the qualitative point of view it is clear that the
 original family 31 must have been much larger than it is today, with
 the majority of the original members now being interstellar.


As for the asymmetry in proper $e$, when the $g-g_6$ resonance is
crossed from higher to lower proper $a$ as in
Figure~\ref{fig:600032_esmot}, there is no ``transport'' along the
resonance but rather there could be a selection effect, by which the
objects exiting $g-g_6$ at high eccentricity are no longer there,
while the ones lucky enough to exit at low to moderate eccentricity
form the IN side of the family.
This IN side of 31 has a proper element span $0.149<e<0.206$ and a
mean of $0.177$; we do not have a quantitative theory explaining these
values, not even approximately, but qualitatively it is possible that
the passage across the resonance $g-g_6$ with $da/dt<0$ leads to lower
proper $e$ for a selected minority of survivors. It is not possible to
test this by a Monte Carlo approach, because it is not appropriate to
select the initial conditions randomly, but it is necessary to take
into account the asymmetry of the ejection velocities and the
correlation between change in orbital elements and the orientation of
the spin axis; this is difficult to model in a quantitative and
reliable way.

The problem which is not solved by the model above is that the number
density as a function of proper $a$ is highly variable, with two peaks
around $a=3.12$ and $3.16$ au. The gap around $a=3.145$ is due to the
$g-g_6$ resonance, not to the YORP eye, which would be located in a
nearly empty region of comparatively large bodies \citep{paolicchiYORP}. 

\begin{figure}[h]
\centering
\figfiginclnocap{11cm}{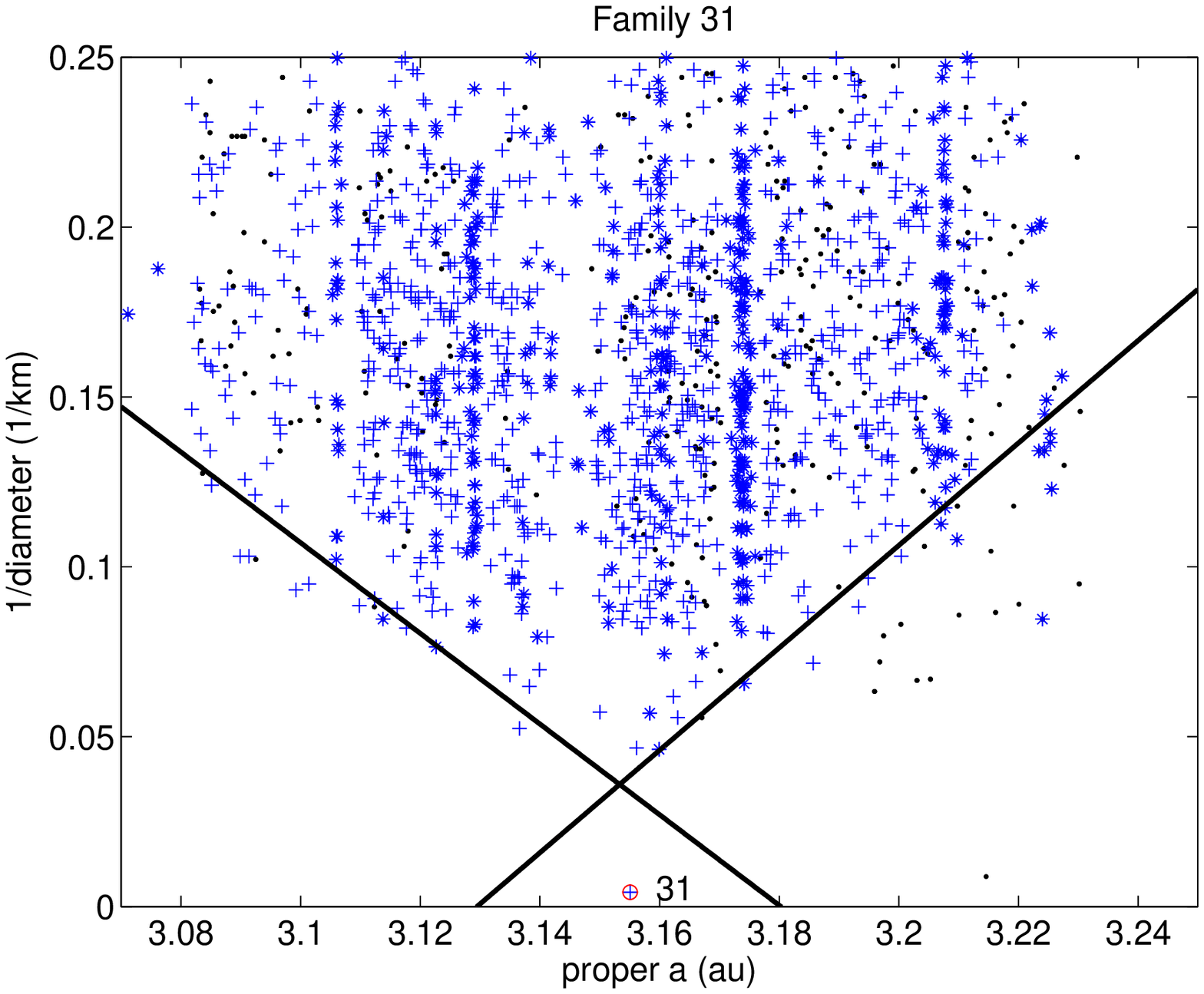} 

\figfigincl{13cm}{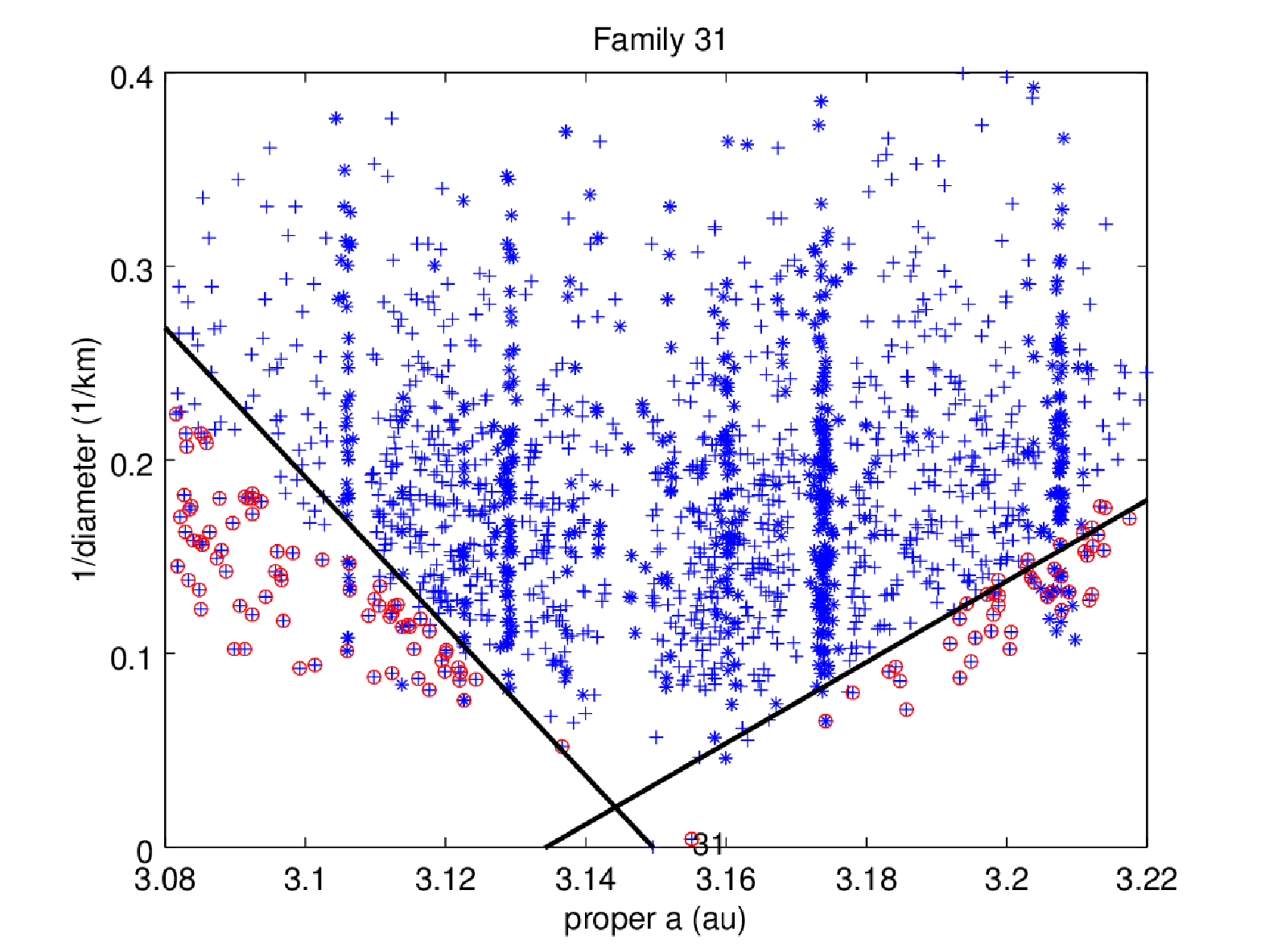}{Top: V-shape for the family of (31)
  Euphrosyne, considered as a single family, showing the negative
  V-base.  Bottom: one possible decomposition of the family 31,
  showing an additional V-shape with two different slopes: this
  implies a model with three separate collisions.}
\end{figure}

A possible way to explain this structure is to assume that the age
computed by the V-shape of Figure~\ref{fig:31_vshapea_outliers} (top)
only refers to the largest remnants of an ancient family, with an age
estimated at $\simeq 1.2$ Gyr, consistent with a YORP age which is
determined by a gap around $D\simeq 14$ km (see Paolicchi et al.,
2018, \textit{in preparation}); to the contrary, the gap occurring
also at much smaller diameters around $a=3.145$ au has nothing to do
with the YORP effect, but with the dynamical removal by the $g-g_6$
resonance.

The alternative explanation is that the concentrations of smaller
bodies belong to more recent collisional families, which in
Figure~\ref{fig:31_vshapea_outliers} (top) are somewhat detached from
the larger members forming the
V-shape. Figure~\ref{fig:31_vshapea_outliers} (bottom) shows one
possible, but by no means unique, decomposition in which smaller
members form a V-shape with higher slopes, and therefore lower ages,
different between the IN and the OUT side. For the OUT side the
inverse slope $1/S=0.478\pm 0.035$, corresponding to an estimated age
of $778\pm165$ Myr, on the IN side $1/S=-0.259\pm 0.007$, corresponding
to an age of $422\pm 85$ Myr. Since the discordance of the ages implies
that these two possible families, one on the IN side, that is
$g-g_6<-2$ of the secular resonance, the other on the OUT side, that
is $g-g_6>+2$ ``/y, could have had different parent bodies, with
different proper $e$. Moreover, the members of the two possible
families currently found would not have crossed the $g-g_6$ resonance:
the ones that had entered the resonance might have been eliminated by
the Swiss roulette.

We acknowledge that we have found two possible interpretations
of the collisional and dynamical history of the dynamical family 31,
without sufficient evidence to select one of the two. 

\subsection{Family of (480) Hansa and of (163) Erigone}

Family 480 is affected by the secular resonance
$2g-g_5-g_6+s-s_6$, with very different effects on the IN
and OUT sides. 

\begin{figure}
\centering
\figfiginclnocap{11.5 cm}{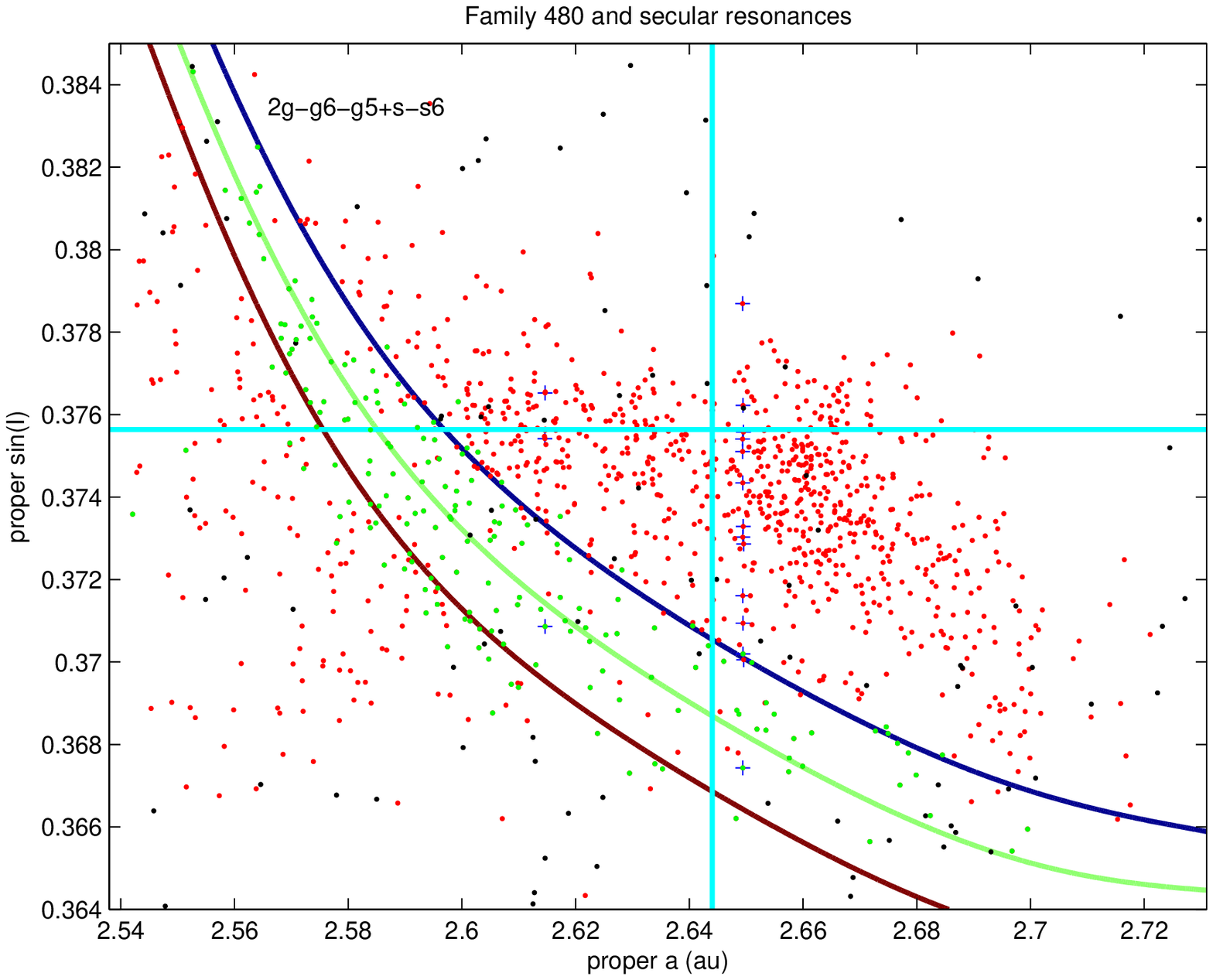}

\figfigincl{11.5 cm}{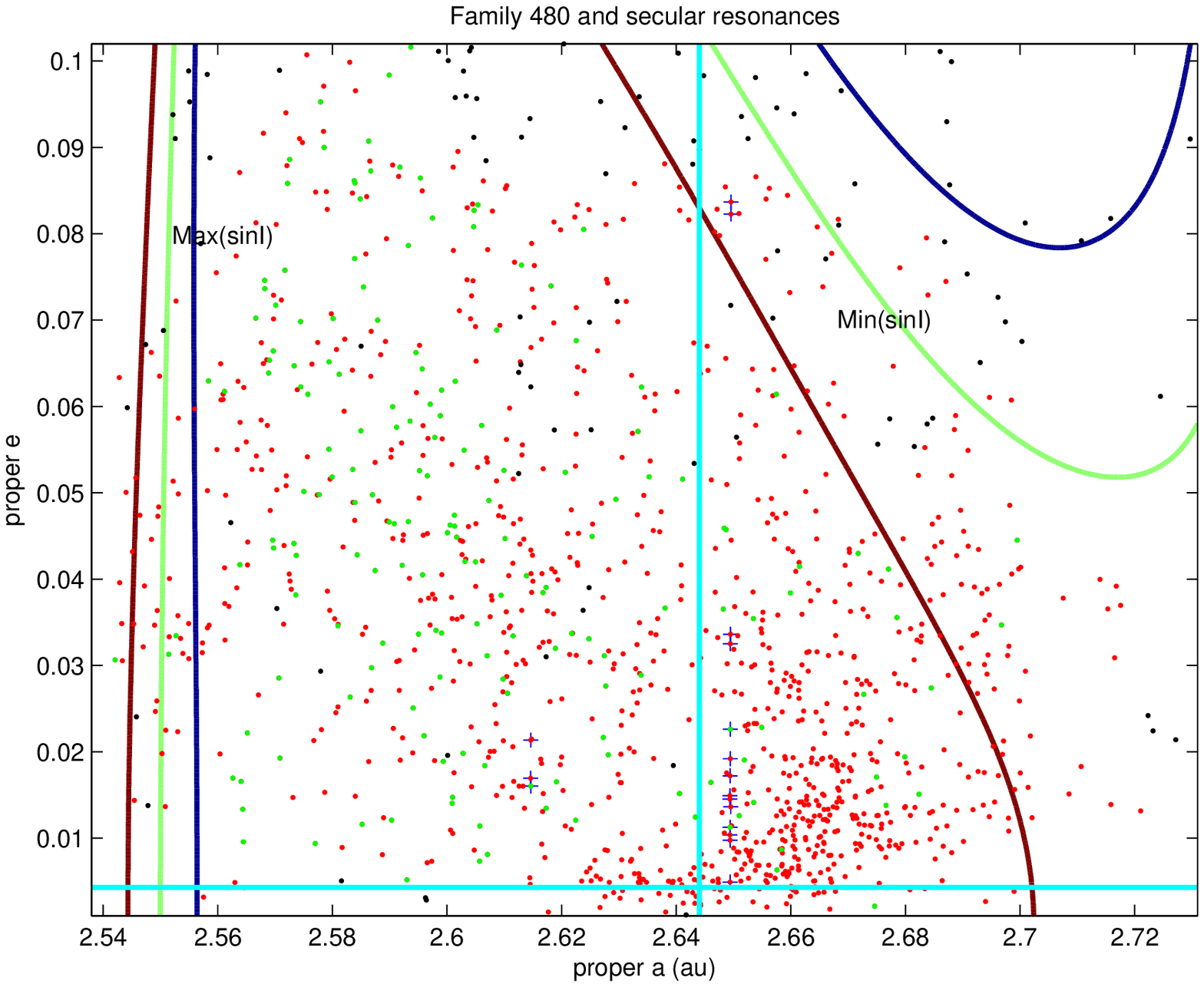}{Projection of the family of (480)
  Hansa on the proper $(a,\sin{I})$ plane (top). Blue crosses indicate
  chaotic members: they are in the $11/4$ resonance with Jupiter. The
  green points indicate that the divisor $2g-g_5-g_6 +s-s_6$ is
  smaller in absolute value than $0.5$ arcsec/y. The same family in
  the proper $(a,e)$ plane (bottom). The level curves of the divisor
  are explained in the text.}
\end{figure}

The increased dispersion of proper $e$ for the portion of the family
with $a<a_0$, where $a_0$ is the proper $a$ of the parent body, shown
in Figure~\ref{fig:480_ae_synt} (bottom) and in the line $480-$ of
Table~\ref{tab:anisotr} can be explained by
Figure~\ref{fig:480_ae_synt} (top) because most of the intersection
between the secular resonance and the family is for $a<a_0$. Moreover,
the members in the IN side are moving towards lower values of $a$ due
to the Yarkovsky effect, and therefore many members currently outside
of the secular resonance zone must have passed through it in the past:
in Figure~\ref{fig:480_ae_synt} (top) these are the points marked in
red but on the left of the strip of blue crosses. We note that the
family members moving towards lower proper $a$ are the ones originally
ejected to an orbit with $a<a_0$, as suggested by the positive V-base
of the V-shape; see Figure~\ref{fig:480_vshapea}.

Also in Figure~\ref{fig:480_ae_synt} we show the level curves
$2g-g_6-g_5+s-s_6=(-0.5, 0, +0.5)$ in arcsec/y, computed by means of
our new synthetic theory for the secular frequencies $g,s$. In the
$(a,\sin{I})$ plane of Figure~\ref{fig:480_ae_synt} (top) the
resonance strip is well defined: although we have computed it only for
the mean value of proper $e$ in the family, which is $0.342$, the
asteroids strongly affected by this resonance are all either within or
very near the strip. On the contrary, in the plane $(a,e)$ of
Figure~\ref{fig:480_ae_synt} (bottom) the resonance strip moves very
much as the third coordinate, proper $\sin{I}$, changes in the family
range, which is $0.364<\sin{I}<0.385$: we are showing the two resonant
strips for the minimum and maximum of $\sin{I}$, demonstrating that
the resonance sweeps the entire range in proper $a$ and $e$. This can
explain the spread of green points in Figure~\ref{fig:480_ae_synt},
(bottom) indicating the members affected by the resonance, both in the
IN and OUT portions of the family.

\begin{figure}[h!]
\centering \figfigincl{10 cm}{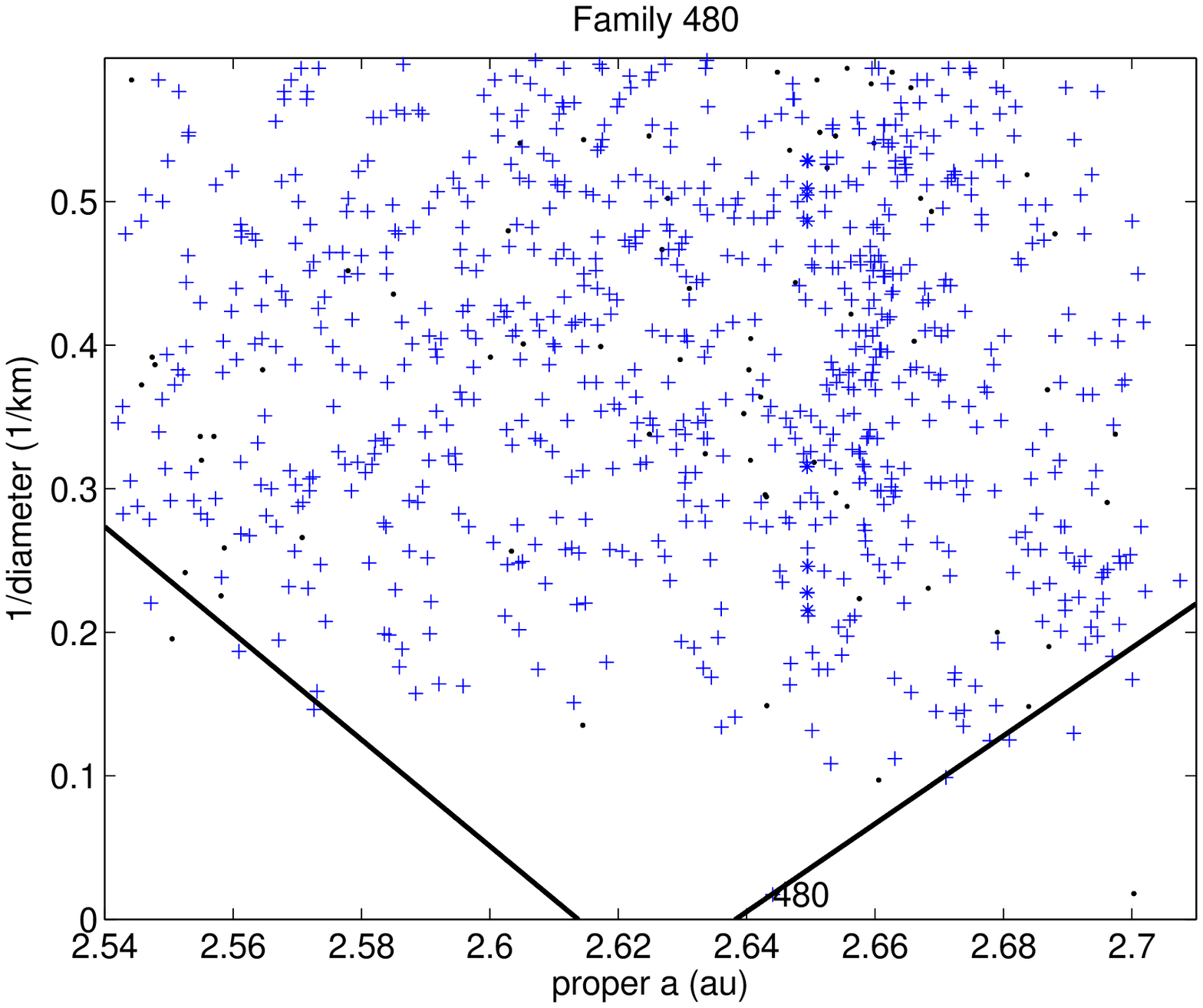}{Family of (480) Hansa
  projected on the plane proper $a$ vs. $1/D$, where the diameter $D$
  has been estimated from the absolute magnitude $H$ and the mean
  family albedo. Note that the V-base is positive.}
\end{figure}

The scattering takes place during the time span in which the family
members are crossing the resonant strip, pushed by Yarkovsky: the
width of the libration strip is narrower than the resonance strip
shown in the figure, but we have checked that there are indeed members
of the family currently in libration. It does not matter how wide the
actual libration strip is, the fact is that it is a barrier which must
be crossed by the members drifting towards lower values of proper
$a$. This affects the proper $e$ much more than the proper $\sin{I}$
because of the D'Alembert rule, by which the perturbation term
associated with the $2g-g_5-g_6+s-s_6$ contains a factor $e^2 \sin{I}
e_5 e_6\sin{I_6}$.  Of the quantities included in this factor,
$\sin{I}$ is large and the others are small. Therefore, in the Hansa
region the derivative with respect to $\sin{I}$ is much smaller than
the derivative with respect to $e$. It follows from the analytical
theory of secular perturbations that the changes in $e$ due to this
resonance, in this region, are much larger than the ones in $\sin{I}$.
This explains why in Table~\ref{tab:anisotr} the mean and standard
deviation of $\delta\sin{I}$ are much smaller than those of $\delta
e$.

In addition to this, it is also necessary to take into account that
some unusual asymmetry in the proper $\delta e$ is caused by the fact
that proper $e$ cannot be negative, by definition. Indeed, a negative
$\delta e$, starting from the value $0.0043$ of (480), can lead to a
negative $e$, which of course only means a positive proper $e$ with a
shift by $\pi$ of the proper longitude of perihelion $\varpi$.

\begin{figure}
\centering
\figfiginclnocap{11.5 cm}{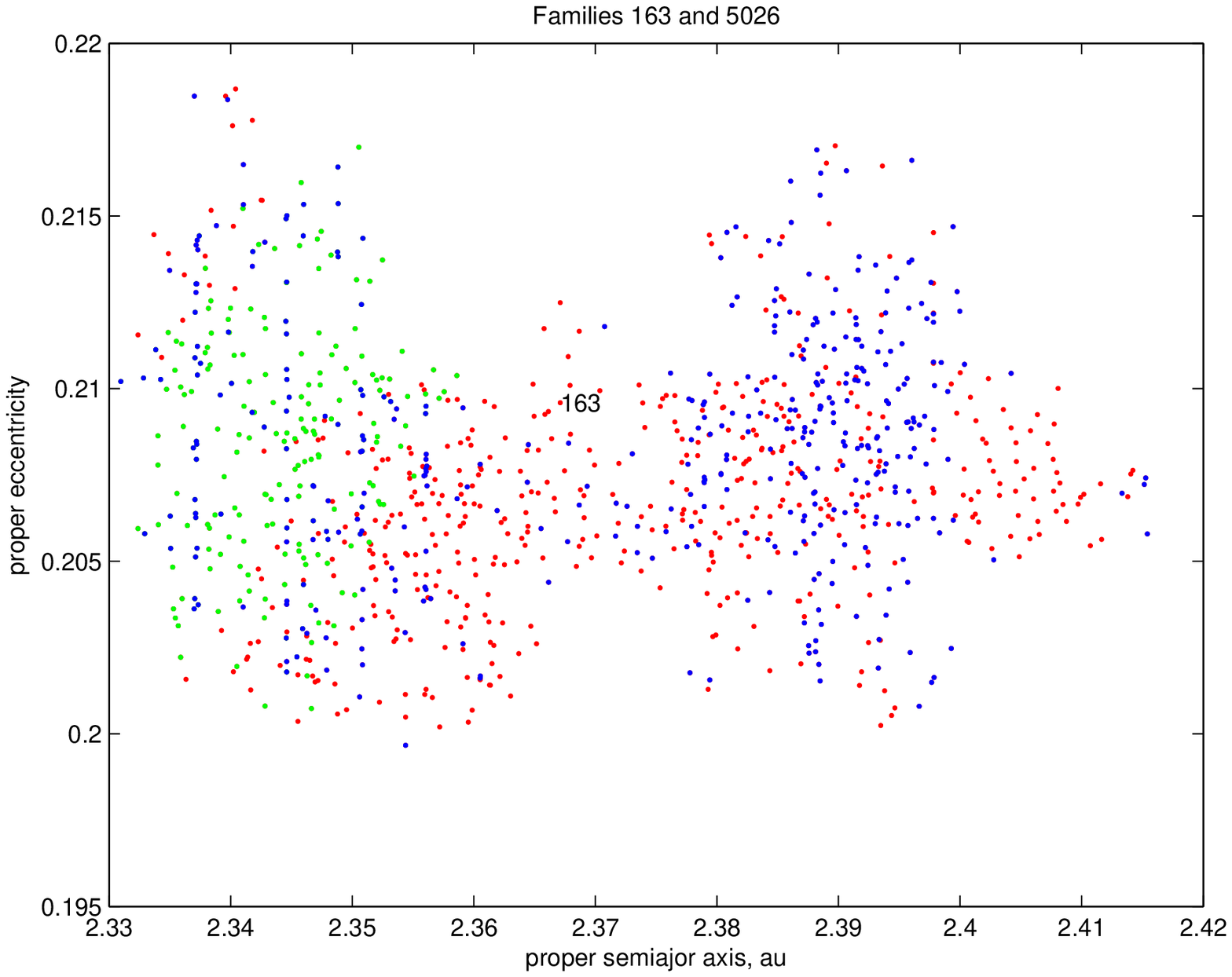}

\figfigincl{11.5 cm}{163_ae}{Projection of the family of (163) Erigone
  on the proper $(a,\sin{I})$ plane (top). Blue crosses indicate
  chaotic members. The green points indicate that the divisor $z_2=2(g-g_6)
  +s-s_6$ is smaller in absolute value than $0.5$ arcsec/y. The same
  family in the proper $(a,e)$ plane (bottom).}
\end{figure}

The family of (163) Erigone, which we have merged with the family
(5026) of Martes, is affected by several mean motion three-body
resonances, resulting in $394$ out of $1023$ members of the merged
family, or $38.5\%$, with Lyapounov time $T_{lyap}<20,000$
yr. Therefore, it is to be expected that the asymmetry is growing with
time because of chaotic diffusion; this applies mostly to the OUT side
of the family and can increase the spread of both $\delta e$ and
$\delta\sin{I}$; see Figure~\ref{fig:163_ae}. However, the asymmetry
is larger in the IN side; and is larger in $\delta\sin{I}$, due to the
fact that the parent body (163) has one of the lowest values of this
proper element.

A larger dynamical effect can be due to the $z_2=2(g-g_6)+s-s_6$
secular resonance, which is very relevant for this family as shown in
\citet{carrubaMNRAS455}. In Figure~\ref{fig:163_ae} the members of the
family with a small divisor $|z_1|=|2(g-g_6)+s-s_6|<0.5$ arcsec/y are
marked in green. The analogy, but also the differences, with
Figure~\ref{fig:480_ae_synt}, are clear: a secular resonance crosses
the family 163 as it is now, but only on the IN side. During the
dynamical evolution of the family, the members with negative Yarkovsky
drift in proper $a$ have been pushed into the secular resonance, and
even beyond it, with the result being an increase of the spread in both
$\delta e$ and $\delta\sin{I}$; the dominant asymmetry is in
$\delta\sin{I}$ because of the position of (163).

\subsection{Family of (179) Klytaemnestra}

The family of (179) Klytaemnestra has for a long time been a problem
in our family classification. Indeed we have not been able to identify
a meaningful V-shape from which to compute its age; family 179
currently has $513$ members, while we have computed ages for all other
families in our classification with $> 300$ members\footnote{Families
  490 and 778 have $>300$ members, but have recent ages ($<10$ Myr)
  which had to be computed with different methods.}.

In addition, from Table~\ref{tab:anisotr} we find a strong asymmetry,
especially in $\delta e$, which in the OUT side is almost five times
the CEV from (179), too much to be accepted as a realistic initial
velocity distribution. We therefore have a dynamical family,
statistically very significant as density contrast in proper element
space, for which we do not have a plausible collisional model. We have
always emphasized that asteroid families are statistical entities,
that is, their membership can never be completely and exactly
identified. However, to explain the bizarre shape of family 179 it is
not enough to remove a few interlopers; we need to consider
decomposing the family into components for which it is possible to
provide collisional and dynamical interpretations.

\begin{figure}[t!]
\centering
\figfigincl{12 cm}{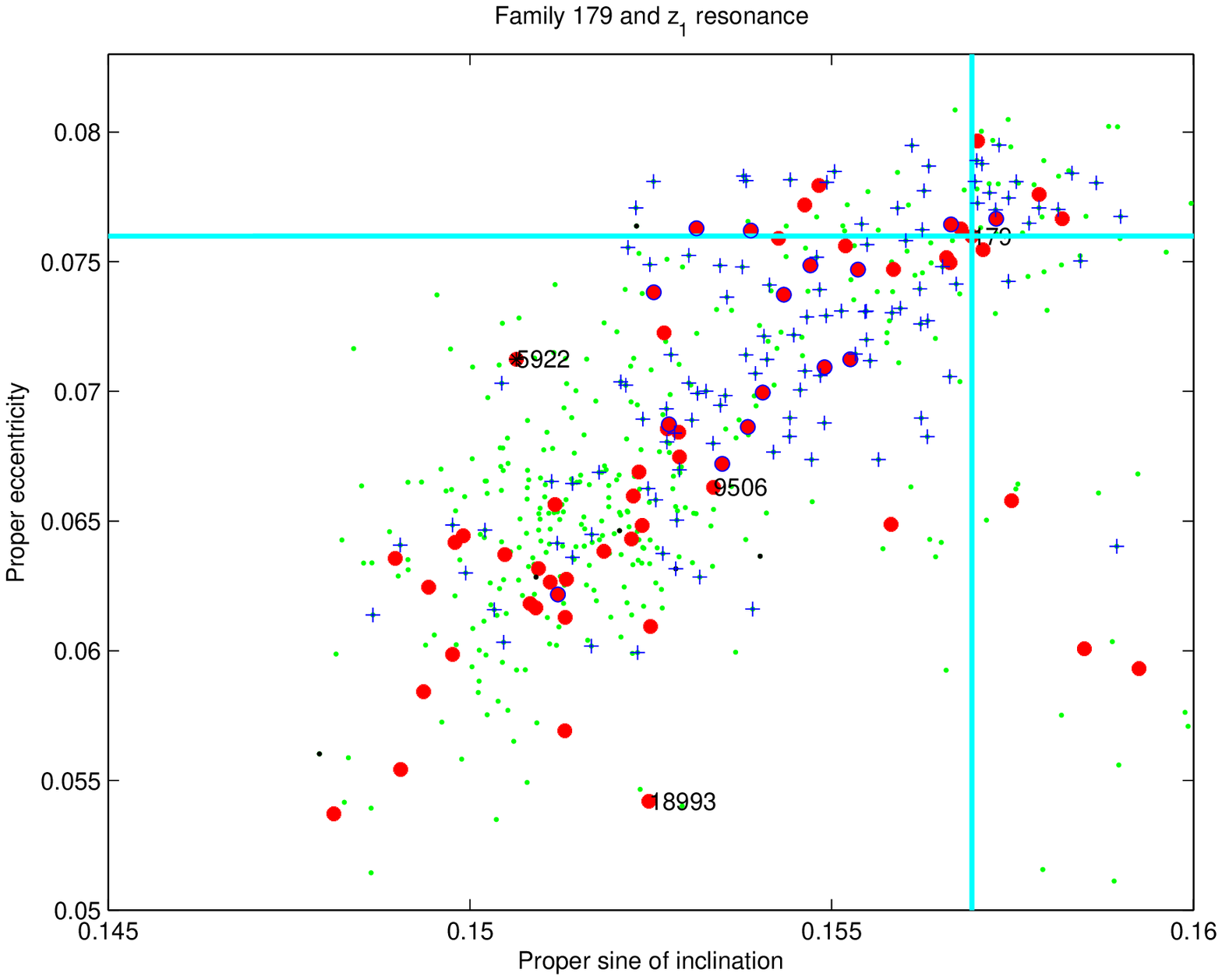}{Dynamical family 179 projected on
  the proper $(\sin{I},e)$ plane. The cyan cross indicates the
  position of the largest member, (179) Klytaemnestra. Most of the
  family members, more than $300$, appear to belong to a clump
  centered too far from (179) for a physically possible distribution
  of ejection velocities. Red stars are core family members (with
  absolute magnitude $H<14$, green points are fainter members attached
  to the core. The members with $|z_1|<0.5$ arcsec/y are marked by
  blue circles for core members and blue crosses for attached ones.}
\end{figure}

The first indication that this might be advisable comes from
Figure~\ref{fig:179_Ie_res}, showing that in the proper $(\sin{I},e)$
plane this family appears as bimodal, with a larger (in number of
members) component far from another, smaller component including the
dominant body (179) Klytaemnestra. Although in Table~\ref{tab:frac}
the family 179 appears with a fraction of fragments $f_v=0.047$, the
global shape of the family is incompatible with a cratering event on
(179): how can fragments from a crater form a compact swarm of
fragments at a distance in velocity space so much larger than the
escape velocity?  The best possible explanation is that this could be
a typical failure of the HCM method, probably in the form of chaining
at the stage of formation of the core family (indicated by red stars
in Figure~\ref{fig:179_Ie_res}), obtained using only proper elements
of asteroids with absolute magnitude $H<14$. Indeed, the fainter
members (green points, $H\geq 14$) attached to the core follow the
elongated shape of the core family; we note that (5922) is a C-type
interloper in an S-type family (marked with a black star; other
attached C interlopers are marked with a black dot). This suggests
that the dynamical family 179 should be decomposed in at least two
collisional families, a smaller one containing (179) and the larger
one with (9506) Telramund as the least populated.

Another element of the explanation we propose is the presence of the
secular resonance $z_1=g-g_6+s-s_6$ which crosses family 179 but
also heavily affects the much larger family 221 \citep{vok}. By marking
in Figure~\ref{fig:179_Ie_res} the family members with $|z_1|<0.5$
arcsec/y, it is clear that they form a bridge connecting the cluster
containing (179), with $z_1<-0.5$ arcsec/y, and the cluster containing
(9506), with $z_1>0.5$ arcsec/y. If these resonant members belong
neither to the collisional family including (9506), nor to the one
including (179), then they are responsible for the chaining.

\begin{figure}[t!]
\centering
\figfigincl{12 cm}{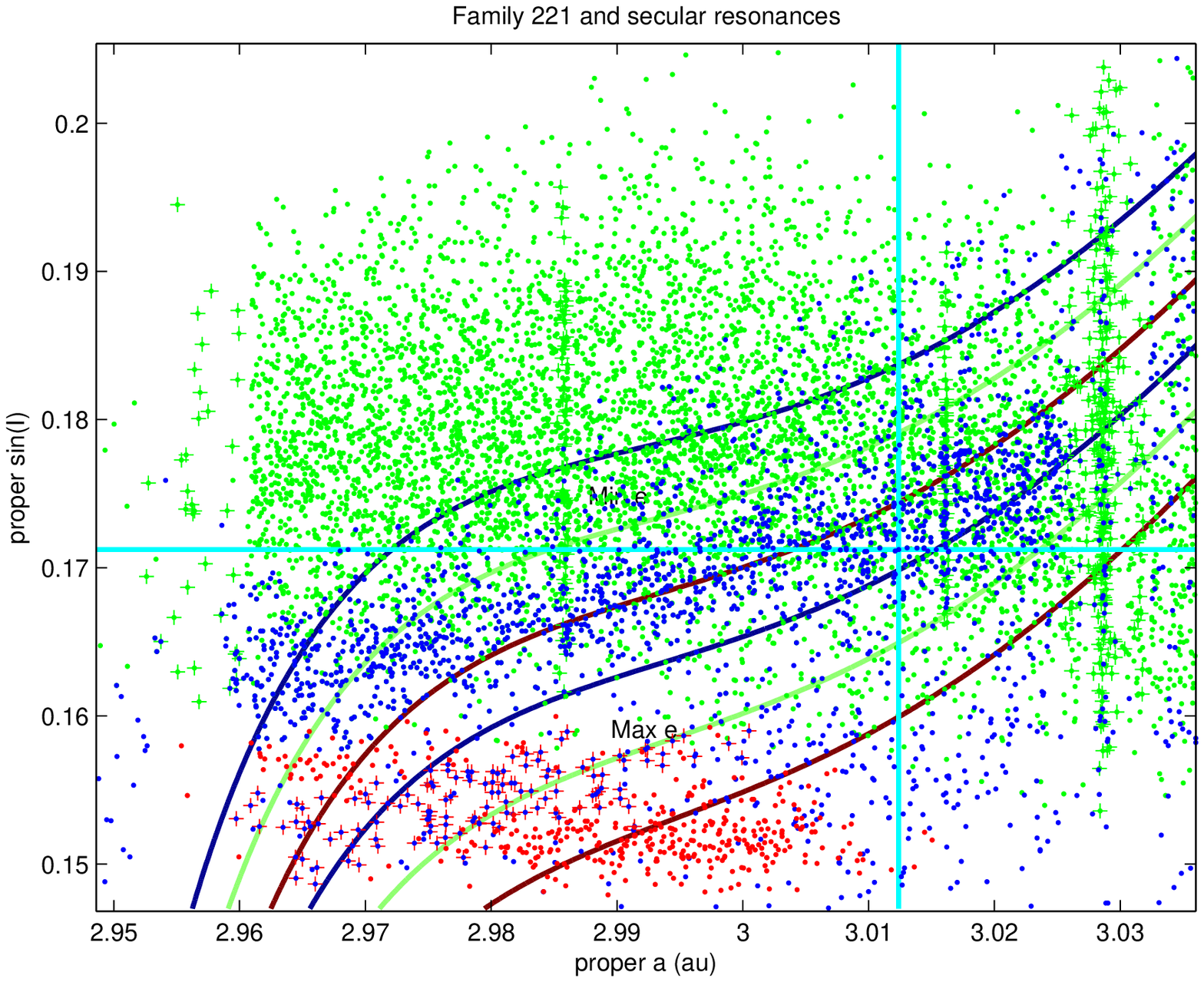}{Projection of the dynamical family
  of (179) Klytaemnestra, and of part of the family of (221) Eos,
  on the proper $(a,\sin{I})$ plane. The members of Eos are in green,
  in blue if they are within $0.5$ arcsec/y from the
  $z_1=g-g_6+s-s_6=0$ secular resonance. The ones of Klytaemnestra are
  red points if not in the resonance, red crosses overlaid by blue
  points if they have $|z_1|<0.5$ arcsec/y. The dynamical
  family 179 is split in two parts by the resonance, which also
  affects 221.}
\end{figure}

Using our new synthetic theory for the secular frequencies $g,s$, and
also the values of the same frequencies, computed together with the
proper elements, for the members of both families 221 and 179, we have
produced Figure~\ref{fig:179_221_z1_aI} showing, in the $(a,\sin{I})$
plane, a portion of the family of (221) Eos and the nearby family
179. We have also overlaid the level lines of $z_1$ computed for two
fixed values of proper $e$, corresponding to the minimum and the
maximum of the values found in family 179 (namely $0.0514$ and
$0.0808$, respectively).

The family of (221) Eos is, in our recently updated classification,
the largest, with more than $16,000$ members. It is clear that such a
family needs to be surrounded by a halo, in which a good fraction of
the asteroids belong to the family, although with a lower number
density than in the recognized family; see
\citet{brozmorby,tsirvoulis}.  The asteroids in the halo have
``escaped'' from the family through different dynamical routes, mostly
driven by the Yarkovsky effect. In particular, if a secular resonance
is effective in a layer in proper element space which has a large
width in proper $a$, a Yarkovsky driven transport is
possible\footnote{We note that the case of resonance $g-g_6$ inside
  the family 31, discussed above, is different because the width in
  proper $a$ of the resonance is small.}. Indeed,
\citet{carrubaMNRAS437} show the results of a large-scale numerical
experiment on transport due to the Yarkovsky effect inside and near
secular resonances, and in particular the $z_1$ resonance; see their
Figure~9, in the top-left corner, which refers to a portion of the Eos
family more or less corresponding to the one shown in our
Figure~\ref{fig:179_221_z1_aI}. The size of the transport in proper
$e$ and $\sin{I}$ shown by these experiments is sufficient to explain
the contamination of the dynamical family 179 with resonant escapers
from Eos.

If we assume that resonant members of 221 can be transported along the
resonance, as driven by the negative $da/dt$ due to the Yarkovsky
effect (as shown by the cyan cross (221) Eos itself is at a larger
value of proper $a$), then the transport is towards lower values of
proper $\sin{I}$. In this way escapers from 221 may enter the region
of the dynamical 179 family, and contribute to the chaining effect
joining the two separate clusters.  An alternative route is possible
(see Figure~\ref{fig:179_221_z1_aI}) by moving to lower proper $a$
until crossing the three-body resonance at proper $a\simeq 2.985$,
then along the mean motion resonance to lower values of proper
$\sin{I}$, before exiting from it inside the secular resonance and
ending up in the region of family 179.

\begin{figure}[h]
\centering
\figfigincl{12 cm}{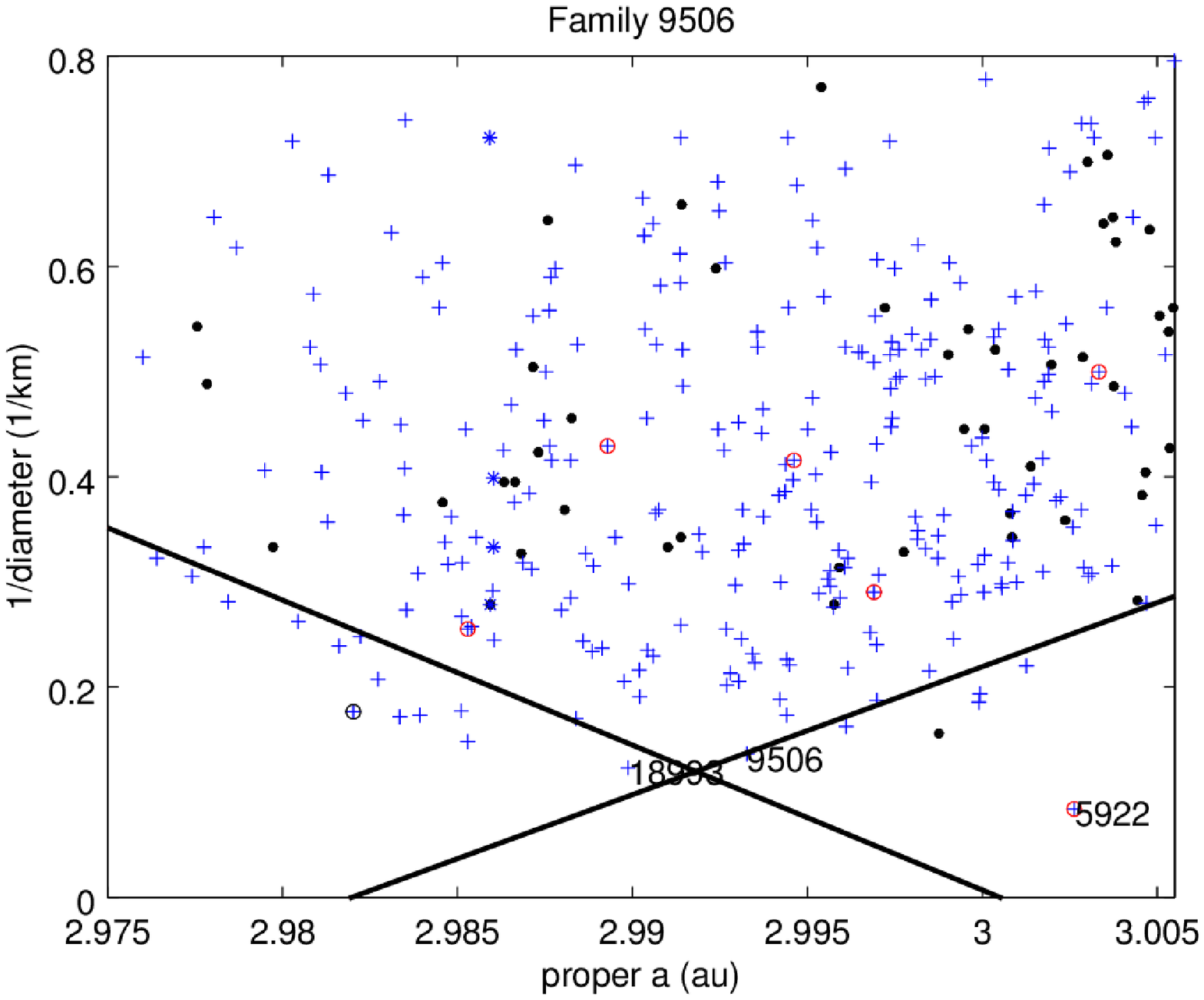}{Proposed family of (9506)
  Telramund, in the $(a, 1/D)$ plane; the C-type interlopers are
  marked with a red circle, the outliers rejected from the fit are
  marked with a blue circle. The two sides of the V-shape have
  compatible slopes.}
\end{figure}  

As suggested by the arguments above, we propose that family 179 should
be split into three parts: a small cluster around (179) with
$z_1<-0.5$, the resonant interlopers ($|z_1|<0.5$ arcsec/y), and a
family currently with $321$ members, (9506) Telramund, with $z_1>+0.5$
arcsec/y\footnote{\citet{nesvast4} propose a Telramund family in their
  Table~2, but leave the doubt that (179) might belong to it. We
  believe we have solved this problem.}.  The smaller cluster around
(179) is too small ($65$ members) to be interpreted, but could be a
cratering family distinct from both 221 and 9506.

The reality of family 9506 can be confirmed by computing a V-shape in
the plane with coordinates proper $a$ and $1/D$, where the diameter
$D$ has been computed from absolute magnitude $H$ assuming an average
albedo (which is $0.25$ after removing the interlopers with albedos of
$<0.1$): this is shown in Figure~\ref{fig:9506_vshapea_numbers}. The
fit gives an inverse slope for the IN side of $-0.073\pm 0.020$, and
$+0.082\pm 0.018$ for the OUT side, that is, they are compatible and
the family has just one age, which is around $220$ Myr (see
Table~\ref{tab:age_crat}). Also this age is compatible with the YORP
age (Paolicchi et al. 2018 \textit{MNRAS, accepted for publication}).

Figure~\ref{fig:9506_vshapea_numbers} also shows that this family is a
complete fragmentation; the two largest members, (9506) and (18993),
are of about the same size, and near the center of the proper $a$
distribution, as determined by the V-shape. As shown in
Figure~\ref{fig:179_Ie_res}, the position of these two largest members
is peripheral in the proper $\sin{I}$ and $e$ distribution, but this
asymmetry cannot be interpreted by the same methods used for cratering
families. 

\section{Conclusions}
\label{s:conclusions}

\subsection*{Definition and identification of cratering families}

We have proposed to use the fraction $f_v$ of the total volume of an
asteroid family consisting of fragments (excluding the largest member)
as a metric to discriminate the cratering from the fragmentation
families.  We have tested the $66$ families in our classification
which currently have more than $100$ members and found a bimodal
distribution: $21$ families with $f_v<1/8$ and $37$ with $f_v>1/2$. In
the middle there are only $8$ families, $4$ with $1/8<f_v<0.22$ and
$4$ with $0.27<f_v<0.40$. Therefore, we selected the value $f_v=1/4$
as a boundary: if $f_v>1/4$ we say that the family is the results of a
\textbf{fragmentation} event($41$ found) and if $f_v<1/4$ we define a
family as the result of a \textbf{cratering} event ($25$ found). We
additionally use, for the few marginal cases, the terms \textbf{heavy
  cratering} for $1/8<f_v<1/4$ and \textbf{marginal fragmentation} for
$1/4<f_v<1/2$. We do not claim that the specific boundary values we
have chosen have a deep geophysical meaning, but simply that they are
appropriate to describe the distribution empirically found for the
quantity $f_v$.

Although the boundary value of more than $100$ members has been chosen
arbitrarily, just as a round number, it appears that indeed $100$
members are enough to discriminate cratering. We have even identified
some smaller families as being of cratering type. For example, the
family of (2) Pallas with only $62$ members, and even some cratering
families for which we do not have a complete list of members, such as
(91) Aegina, (429) Diotima, and (179) Klytaemnestra, because of
overlap with other families.

An important result is that all these families appear compositionally
homogeneous, in that the number of interlopers (identified by physical
observations, mostly WISE albedos) is small, with an exception being
family 5, for which the presence of another family of asteroids with
incompatible composition had already been proposed in \citet{paperV}.
Nevertheless, we have discarded two namesake asteroids as interlopers:
(110) Lydia and (194) Prokne, and correspondingly adopted the new
namesakes (363) Padua and (686) Gersuind. This is not a surprise,
because the size distribution of the background needs to be more
shallow than that of the family. Indeed, the position inside the
resonance of families with Padua and Gersuind as namesake have already
been proposed in \citet{carrubaMNRAS395} and in
\citet{gilhutton,carrubaMNRAS408,Novakovic2011}, respectively;
moreover, that (110) Lydia undergoes high amplitude libration inside
the $z_1$ resonance has already been shown by~\citet{milkne1992}.

\subsection*{No problem with asymmetry for most cratering families}

We then analyzed the scatter of the proper elements in the $(e,
\sin{I})$ plane with respect to the parent body to check whether
it is compatible with a realistic model of the relative velocities of
the fragments, immediately after ejection from the gravitational
sphere of influence of the parent body.

In a way, the most important result shown by the summary
Table~\ref{tab:anisotr} is that, in most cases, there is nothing
remarkable in the four values of the mean and standard deviation for
both $\delta e$ and $\delta \sin{I}$.  Indeed, in $15$ out of $25$
cratering families (with more than $100$ members) there appears to be
no problem, that is, the first two moments of the distributions of
both proper $e$ and $\sin{I}$ are of the order of the CEV. This
certainly occurs for families 3, 5, 10, 686, 302, 396, 606, 363, 1303,
1547, 87, 148, 778.

The following five families have asymmetries explained by the presence
(either known or at least proposed) of two collisional families: 4,
15, 283, 20, 569. For three families we believe the number of members is
currently too low to draw reliable conclusions: 1222, 96, 410.

\subsection*{Difficult cases, only partially explained}

This left us with only $4$ families out of $25$ for which we had to
look for an explanation for an anomalous asymmetry: 31, 179, 480, and
163. In these $4$ cases the asymmetry is too large to be attributed to
the initial escape velocity of the fragments from a cratering event.
For all these we have found at least a plausible cause for the
anomalous asymmetries.

For the family (31) Euphrosyne we have a dynamical explanation, based
on the negative V-shape and the consequent Yarkovsky evolution which
leads to a past crossing of the $g-g_6$ secular resonance with either
increasing or decreasing proper $a$. This model, together with the
suggestions from some numerical-propagation experiments, proposes that
a large fraction of the original fragments has been ejected from the
family, mostly on hyperbolic orbits. An alternative model assumes that
the original collisional family, which has an age of $> 1$ Gyr old,
has been deeply eroded (see \citet{paperV}) by the same dynamical
mechanism as in the other, above model, but the central part of the
family has been replenished by a number of more recent collisions;
probably two of them, one for each side with respect to the $g-g_6$
resonance. We do not have enough information to decide which of these
two models represents the true history of family 31, and therefore it
is not even clear if the main cause is collisional rather than
dynamical.

For the family of (480) Hansa we have identified the secular resonance
$2g-g_5-g_6+s-s_6$ , combined with the effects of the Yarkovsky
secular perturbation, as the main cause of the scattering of the
proper $e$ to values significantly higher than those of (31)
itself. Some contribution to the asymmetry parameters for $\delta e$
could also be due to the fact that proper $e$ is greater than $0$ by
definition. In this case the explanation is fully dynamical. For the
family of (163) Erigone we also find a dynamical explanation for the
current family shape, that is, due to the consequences of both a
secular resonance and several mean-motion resonances.

For the family of (179) Klytaemnestra, which had been rated as a
problem for its unexplained shape in our previous papers
\citep{paperIII, paperIV, paperV}, we have found an explanation by
decomposing the dynamical family, as assembled by our multistage HCM
procedure, into three pieces, each of a different origin. We propose
that one component is composed by escapers from the extra large family
of (221) Eos, transported also by means of the $z_1=g-g_6+s-s_6=0$;
the second is a small cratering family from (179), and the third (with
most members) a fragmentation family with namesake (9506)
Telramund. This model has been confirmed by finding a good age
estimate, using our V-shape method, for family 9506. In this case the
explanation requires both dynamics and a different collisional model,
with two collisions.

All these three solutions of the problem of a realistic collisional
model need to be confirmed by additional work, both by dynamical
studies and using new and improved data.  We confirm that the
existence of a realistic collisional model for every single dynamical
family cannot be taken for granted. Indeed, in one case we now think
it does not exist, and the dynamical family 179 needs to be
reinterpreted with a completely different collisional model,
and with a largely different membership with respect to the one
suggested by the HCM method.

The fact that this was found necessary in only $1$ out of $25$
cratering families, with $2$ more dubious cases of families with $\leq
120$ members, indicates that the HCM method is not bad at all, but
nonetheless must not be taken as a ground truth. In
\citet{paperV}[Section 5] we already indicated a few fragmentation
families for which to obtain a consistent collisional and dynamical
explanation we need to split dynamical families into three, four, or
even more components. In another case, family 163, we had already
proposed to merge two dynamical families into one collisional family,
and this choice has been supported by the asymmetry data.

We believe that this paper has shown that the dynamical families,
obtained by an automated HCM procedure in the space of proper
elements, can provide information not only on the existence of either
one or more ``true'' collisional families in them, but also
first-order information on the original distribution of ejection
velocities of the fragments, in the case of cratering events. If there
are some cases in which there are problems, we can identify them, and
find reasonable dynamical explanations in most cases. In a few cases,
the outcome of the HCM procedure needs to be modified, and this can
also be done in a rational way.

\subsection*{Possible future work}

The best way to improve on the understanding of cratering families is
to obtain more data, in particular more members for the families
currently in the range between $100$ and $200$ members, and more
physical data to perform a much better identification of
interlopers. Dynamical studies need to be conducted, in particular on
the problem of Yarkovsky transport along secular resonances, which
obviously depends upon the orientation in the proper element space of
the resonance surface.

For fragmentation families, asymmetry parameters might have
to be defined and used in a different way.

\section*{Acknowledgments}

We thank the referee (V. Carruba) for useful suggestions, in
particular in solving the problems with the two difficult cases of
families 163 and 179.  ZK acknowledges support from the Serbian
Academy of Sciences and Arts via project F187, and from the Ministry of
Education, Science and Technological Development of Serbia through the
project 176011.  PP acknowledges funding by the University of Pisa and
INFN/TASP.

\bibliographystyle{aa}
\bibliography{cratering_ref}

\nobreak
\appendix
\section{Effects of spin-velocity correlation}
\label{s:paol}

The overall properties of ejecta from cratering impacts have been widely 
analyzed in the literature. Most relevant results have been summarized and 
discussed in \citet{Housen_and_Holsapple_2011}. The general features of a 
cratering process can be represented in Figure \ref{fig:cra}.
\begin{figure}[h]
\centering
\figfigincl{10 cm}{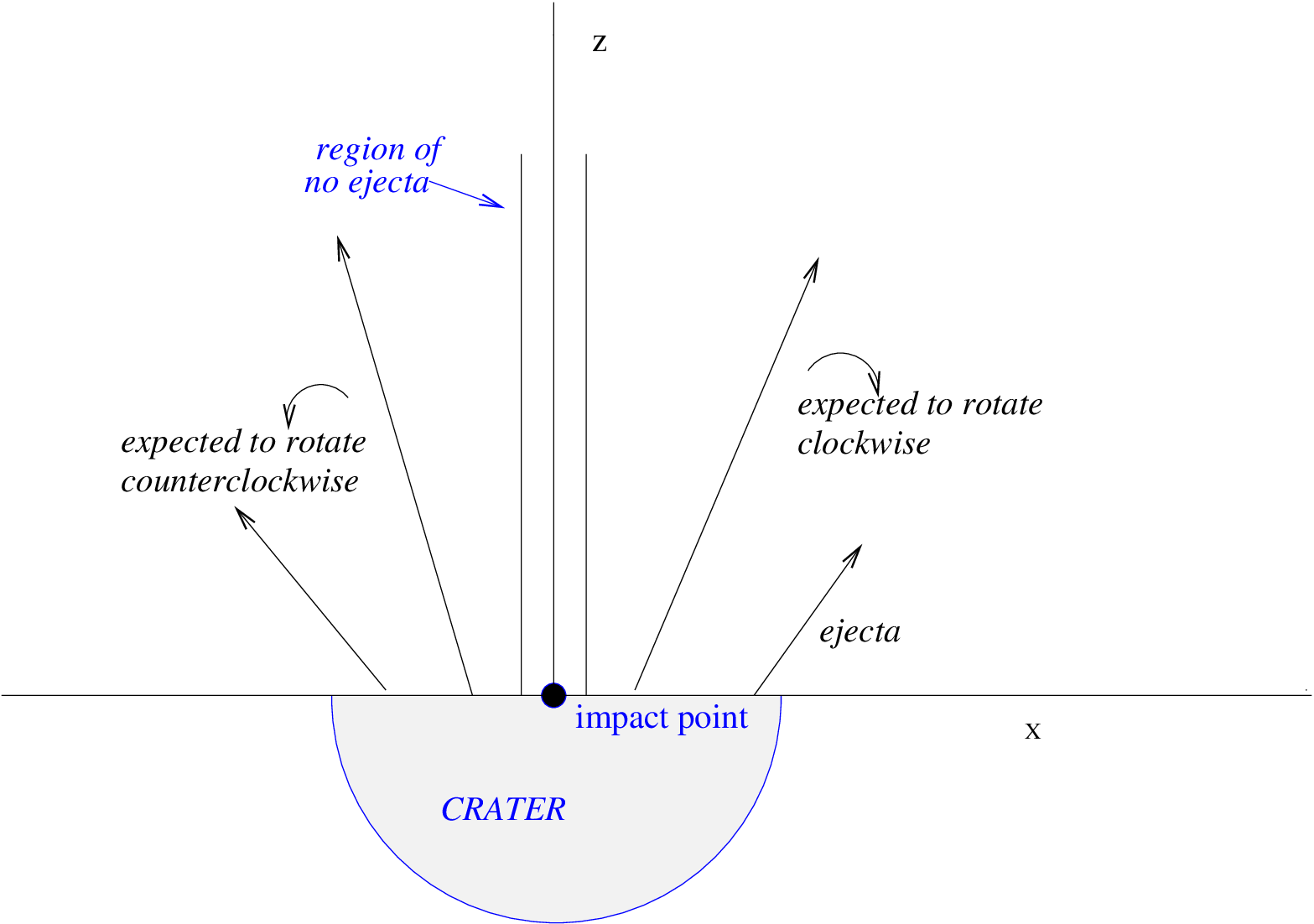}{General features of ejecta from a cratering impact.}
\end{figure}
In this latter quoted publication it is also shown that several results from the experiments 
can be framed within the general scaling equations. For the present task, the 
most relevant equation correlates the velocity of the ejecta ($v$) with the 
distance ($x$) from the impact point:  
\begin{equation}
  v= const \  x^{-1/\mu}  
\end{equation}
\noindent where $\mu$ depends on the properties of the colliding
bodies, but is often $\mu \simeq 0.4$, therefore $v \simeq const
\ x^{-2.5}$.

Unfortunately, in the literature there is no explicit analysis of the
rotational properties of the ejecta from a cratering event.  The
rotational properties of ejecta created in a catastrophic impact have
been widely discussed in the literature \citep{Fujiwara_etal_1989,
  Giblin_etal_1994,Holsapple_etal_2002}.  In
\cite{Fujiwara_and_Tsukamoto_1981} also the properties of the spin
vector have been discussed, showing that the direction of rotation is
correlated with the place of ejection, as shown in the figure.

This experimental evidence, confirmed also by \citet{Giblin_etal_1994}
was included in the so-called semi-empirical model
\citep{Paolicchi_etal_1989, Paolicchi_etal_1996}. In this model the
fragmentation and ejection of fragments are driven by a
\textit{velocity field} $\vec{v(x)}$.  The rotation of fragments due
to the fragmentation process, apart from a term connected to the
shape, is proportional to the rotor $\nabla \times \vec{v}$.  The
dependence of both translational and rotational properties of ejecta
on the residual stress in targets has been discussed by
\citet{Kadono_etal_2009}.

Although the generalization of these ideas to cratering processes is
not based on any experimental evidence, there are several similarities
concerning the ejecta properties, and it is reasonable to assume that
they hold also for cratering.  According to this assumption, the
$v(x)$ relation given above implies also that the
  fragments from a crater should rotate in such a way that the side of
  the body nearest to the impact point rotates away from it, that is
  clockwise on the right of the figure and counterclockwise on the
  left \citet{bigfamilies}[Section 5.2]. In Figure \ref{fig:cra} this
  property is indicated by the labels "expected to rotate
  clockwise/counterclockwise".

If so, also when a family is formed as the outcome of a cratering
event, the Yarkovsky drift may be, depending on the impact geometry,
either parallel or antiparallel to the original $\Delta a$, due to the
ejection velocity, as proposed in \citet{bigfamilies} where
Figure~8 actually refers to a fragmentation family.

A consequence of this effect is the possibility of explaining at least
some negative V-bases. If the fragments ejected at larger $a$ have a
retrograde rotation, their $a$, initially larger, decreases with time
due to Yarkovsky effect. Therefore, original $\delta a$ and Yarkovsky
$da/dt$ have a different sign, and the wings appear to cross at $1/D
>0$ (negative V-base). The family of (31) Euphrosyne discussed in
Section~\ref{s:shapes} exhibits this feature; see
Figure~\ref{fig:31_vshapea_outliers} (top). Moreover, (31) is located
near the outer edge of the most populated portion of the main belt,
implying that a projectile coming from inside (sunward direction) is
more likely; this would result in a negative correlation of the
original $\delta a$ due to the relative velocity of the fragments and
the $da/dt$ from Yarkovsky.

\section{Age estimation for cratering families}
\label{s:age}

This section contains all the ages we have been able to compute for
cratering-type families identified in this paper. Most of these ages
have originally been published in the papers by
\citet{paperIII,paperIV,paperV}; however, in these papers some of
these families were considered as fragmentations, but now we have
found them to be of cratering type, according to our
definition. Moreover, some families had a different namesake (because
we have recognized the previous namesake as an interloper).

Two new ages have been computed in this paper, namely for the family
of (87) Sylvia and (9506) Telramund; however, 9506 is not a cratering
family, as a result of the decomposition of the family 179. Therefore
in Tables~\ref{tab:slope_crat} and \ref{tab:age_crat} we list the age
data of 9506 below a line at the bottom, together with the age for
family 15124 which has been shown to be a fragmentation, although it
was discovered as a subfamily of a cratering-type family.
\begin{figure}[h]
\figfigincl{10 cm}{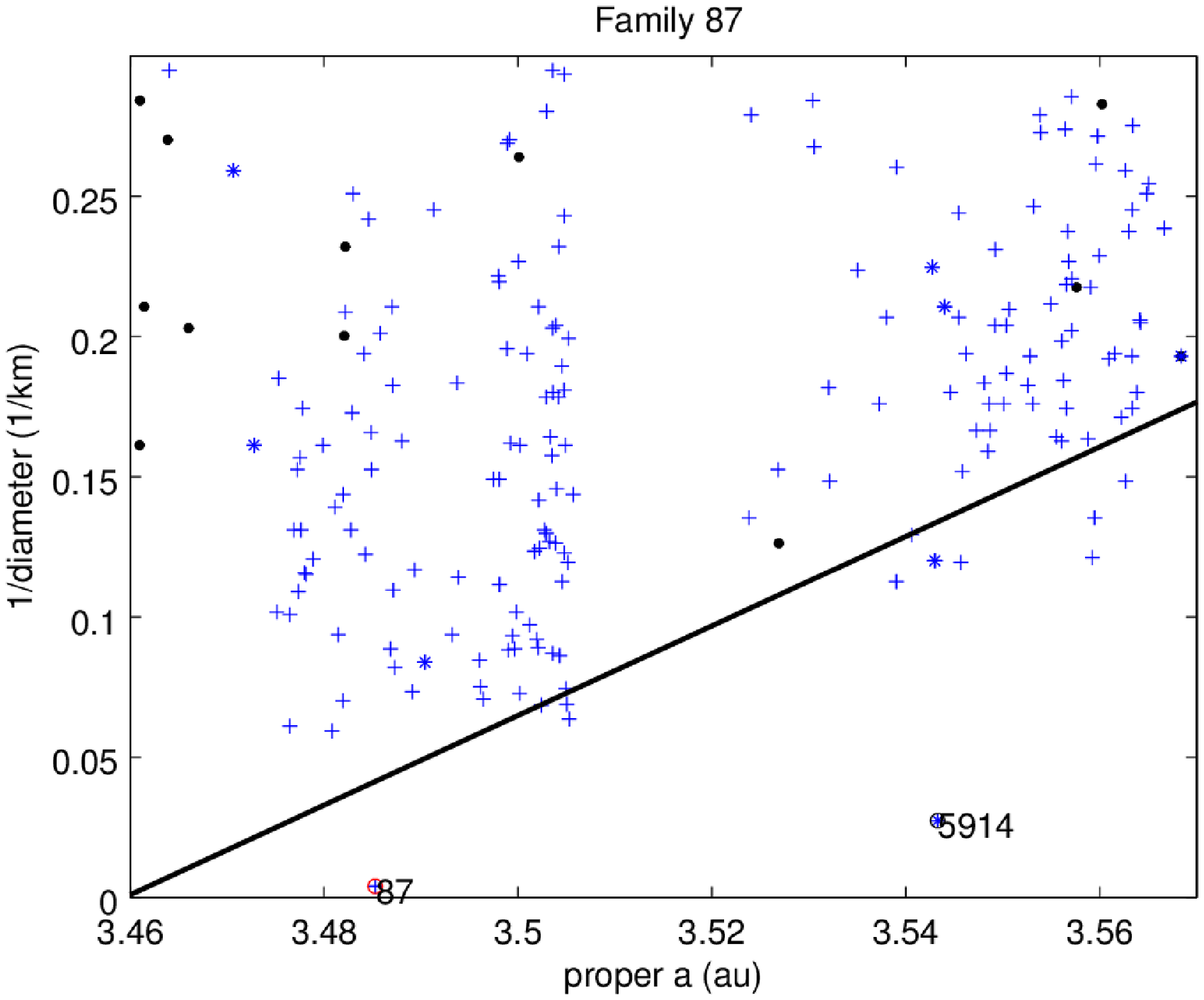}{Family of (87) Sylvia
  on the proper $a$, $1/D$ plane, showing a one-sided
  V-shape, with the gaps due to resonances: (5914) is an outlier.}
\end{figure}
The V-shape for family 87 is shown in Figure~\ref{fig:87_vshapea}: we
note that the IN side of the V-shape is missing, probably because of
the effect of the strong $11/6$ mean motion resonance with Jupiter at
$3.472$ au. The gap in the middle of the remaining OUT side is due to
the resonance $9/5$ with Jupiter at $3.515$ au, therefore it is not
due to the YORP effect. An age of $1220\pm 40$ Myr has been estimated
in \citet{carrubaMNRAS451}; the nominal value is well consistent with
our estimate, while we are rather skeptical about their claim for such
a low uncertainty, because they might not include the calibration
uncertainty. On the contrary, we give a separate estimate for this
error term, which is the dominant one. In a Monte Carlo simulation of
the family formation, the calibration uncertainty is hidden in the
choices made to include the Yarkovsky effect in the numerical
integrations.

The V-shape for the new family of (9506) Telramund was already shown in
Figure~\ref{fig:9506_vshapea_numbers}. 

One of the age estimates merits additional comments to what has
already been written in the previous papers. In \cite{paperIII} we
mentioned that the very young age estimated for the family of (1547)
Nele, as obtained from the V-shape, could be overestimated because of
the contribution of the initial velocity field to the inverse
slope. Recently. \citet{carrubaMNRAS477} provided an estimate of the
age for the same family at about $7$ Myr, calculated using a method
based on the past clustering of the secular arguments $\varpi$ and
$\Omega$, which is not significantly affected by the initial velocity
spread. This implies that of the $14$ Myr age estimated from the
V-shape about half is due to the contribution from the original
velocity spread. This result is remarkable, because it allows, by
scaling (linearly with the diameter of the parent body), to estimate
the order of magnitude of this contribution for other families. As an
example, for the family of (31) Euphrosyne the age estimated from the
main V-shape (Figure~\ref{fig:31_vshapea_outliers}), about $1,400\pm
300$ Myr, is affected by a contribution from the initial velocity
spread of $\sim 100$ Myr, which is less than the uncertainty of the age
estimate, implying that it is not strictly necessary to include this
contribution. We note that it can be both positive and negative (see
Section~\ref{s:shapes}), and this may contribute to explain why the
method used in \cite{carrubaAPJ2014} estimates a lower age than for
family 31 than our method does here: their family evolution simulation
assumes an isotropic velocity field, implying that the contribution to
the inverse slope (and to the age) from the initial velocity spread is
positive.

Table~\ref{tab:slope_crat} contains the data on the fits of V-shapes
in the $(a, 1/D)$ plane: family number/name, number of members, side,
slope ($S$), inverse slope ($1/S$), STD of $1/S$, ratio OUT/IN of
$1/S$, and STD of the ratio.

Table~\ref{tab:age_crat} gives the age estimation for the cratering
families: family number and name, $da/dt$, age estimation, uncertainty
of the age due to the fit, uncertainty of the age due to the
calibration, and total uncertainty of the age estimation.

The Tables~3 and 4 do not contain the data on the new supposed
subfamilies of family 31, because the necessary family split has not
been identified unambiguously. However, in the graphic summary of all
the ages of cratering families, Figure~\ref{fig:famage_plot_31_87}, we
have also indicated the two additional ages which could be found in
family 31, to show that there would be nothing strange in assuming
that a parent body as large as (31) Euphrosyne could have been
affected by multiple craterings, spaced several hundreds of millions
of years apart.

\nobreak
\begin{figure}[h]
\centering
\figfigincl{11 cm}{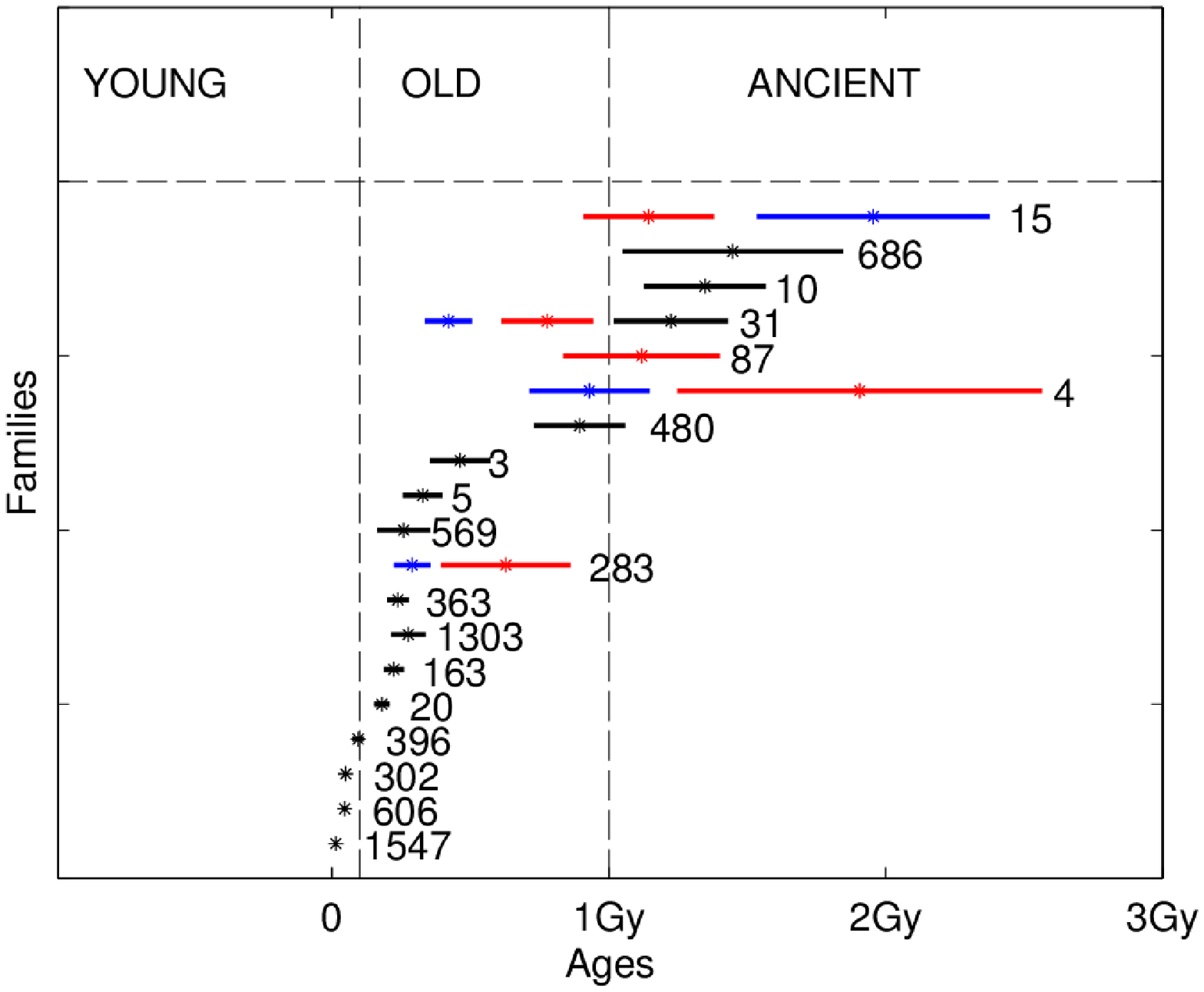}{Ages of cratering families;
  colored bars are for families with two ages (red: OUT side, blue:
  IN. Note that family 31 could have three ages.}
\end{figure}

\begin{table}[t!]
\footnotesize
 \centering
 \caption{Slopes of the V-shape for the cratering families.}
  \label{tab:slope_crat}
\medskip
  \begin{tabular}{|lr|crrl|ll|}
  \hline
number/ & no.    & side   & S & $1/S$ & STD  & ratio & STD    \\
name    & members&        &   &       &$1/S$ &       & ratio  \\
\hline
3 Juno         &960  & IN & -5.261  & -0.190  & 0.038 &&\\
               &     & OUT&  7.931  &  0.126  & 0.049 &0.66&0.29\\
4 Vesta        &8620 & IN & -2.983  & -0.335  & 0.040 &&\\
               &     & OUT&  1.504  &  0.665  & 0.187 &1.98&0.61\\
5 Astraea      &5192 & IN &  -6.596 & -0.152  & 0.095 &        &\\
               &     & OUT&  -6.845 & -0.146  & 0.017 &0.96& 0.61\\
10 Hygiea      &2615 & IN & -1.327  & -0.754  & 0.079 &&\\
               &     & OUT&  1.329  &  0.752  & 0.101 &1.00&0.17\\
15 Eunomia     &7476 & IN & -1.398  & -0.715  & 0.057 &&\\
               &     & OUT&  2.464  &  0.406  & 0.020 &0.57&0.05\\
20 Massalia    &5510 & IN &-15.062  & -0.066  & 0.003 &&\\
               &     & OUT& 14.162  &  0.071  & 0.006 &1.06&0.10\\
31 Euphrosyne  &1137 & IN & -1.338  & -0.747  & 0.096 &&\\
               &     & OUT&  1.507  &  0.663  & 0.081 &0.89&0.16\\
87 Sylvia      &191  & OUT&  1.597  &  0.626  & 0.096 &&         \\     
163 Erigone    &429  & IN & -7.045  & -0.142  & 0.035 &&\\
               &380  & OUT&  6.553  &  0.153  & 0.013 &1.08&0.28\\
283 Emma       &536  & IN &  -6.046 & -0.165  & 0.019 &    &\\
               &     & OUT&  - 2.814& -0.355  & 0.112 &2.15& 0.72 \\
569 Misa       &441  & IN & -5.0376 & -0.199  & 0.151 &&\\
               &    & OUT&  6.5380 &  0.153 & 0.052 &0.77&0.64\\
686 Gersuind  &       & IN   & -1.758 & -0.569  & 0.322 &        &      \\
              &       & OUT  & 1.874 & 0.534  & 0.138 &  0.94  & 0.58 \\
\hline
302 Clarissa   &222  & IN &-27.170 & -0.037 & 0.007 &    &    \\ 
               &     & OUT& 33.6409&  0.030 & 0.005 &0.81&0.16\\
396 Aeolia     &306  & IN & -32.358  & -0.031  & 0.005 &&\\
               &     & OUT&  35.556  &  0.028  & 0.005 &0.91&0.22\\
606 Brangane   &192  & IN & -54.027  & -0.019  & 0.002 &&\\
               &    & OUT&  60.374  &  0.017  & 0.003 &0.89&0.17\\
\hline
363 Padua      &      & IN   & -7.577 & -0.132  & 0.014 &    &   \\
              &       & OUT  &  8.521 &  0.117  & 0.125 &  0.89  & 0.13 \\
480 Hansa       &960  & IN & -3.710  & -0.270 & 0.109 &    &\\
                &     & OUT&  3.064  &  0.326 & 0.040 &1.21&0.51\\
1303 Luthera   &   251& IN   & -6.465 & -0.155 & 0.014 &    &    \\
               &      & OUT  &  6.633 &  0.151 & 0.019 &0.97&0.15\\
1547 Nele       &152 & IN &-201.336  & -0.005  & 0.0008 &   &\\
                &    & OUT& 187.826  &  0.005  & 0.002 & 1.07 & 0.44\\ 
\hline
9506 Telramund & IN & -13.780 & -0.073 & 0.020&     && \\
               & OUT&  12.167 &  0.082 & 0.018& 1.12& 0.40& \\
15124 2000EZ$_{39}$&  & IN & -14.422 & -0.069 & 0.006 &    &\\
                &    & OUT&  14.337 &  0.070 & 0.007 &1.01&0.14\\
\hline
\end{tabular}
\end{table}

\newpage

\begin{table}[t!]
\footnotesize
 \centering
 \caption{Age estimation for the cratering families.}
  \label{tab:age_crat}
  \medskip
  \begin{tabular}{|l|cc|rccc|}
  \hline
number/           &side  & $da/dt$        & Age & STD(fit)& STD(cal)&STD(age)\\ 
name              &IN/OUT& $10^{-10} au/d$ & Myr  &   Myr    &   Myr    &   Myr  \\
\hline
3 Juno            &  IN  &    -3.46       &  550&    110   & 110 &   156    \\
                  &  OUT &\phantom{-}3.41 &  370&    143   &  74 &   161    \\
4 Vesta           &  IN  &    -3.60       &  930&    112   & 186 &   217    \\ 
                  &  OUT &\phantom{-}3.49 & 1906&    537   & 381 &   659    \\
5 Astraea         &  IN  &    -3.72       &  408&    256   &  82 &   269    \\ 
                  &  OUT &\phantom{-}3.70 & 395 &     45   &  79 &    91    \\ 
10 Hygiea         &  IN  &    -5.67       & 1330&    139   & 266 &   300    \\
                  &  OUT &\phantom{-}5.50 & 1368&    183   & 274 &   329    \\
15 Eunomia        &  IN  &    -3.66       & 1955&    155   & 391 &   421    \\
                  &  OUT &\phantom{-}3.55 & 1144&     57   & 229 &   236    \\
20 Massalia       &  IN  &    -3.81       &  174&      7   &  35 &    35    \\
                  &  OUT &\phantom{-}3.73 &  189&     16   &  38 &    41    \\
31 Euphrosyne     &  IN  &    -5.71       & 1309&    169   & 262 &   312    \\
                  &  OUT &\phantom{-}5.72 & 1160&    142   & 232 &   272    \\
87 Sylvia         &  OUT &\phantom{-}5.59 & 1120&    172   & 224 &   282    \\
163 Erigone       &  IN  &    -6.68       &  212&     53   &  42 &    68    \\
                  &  OUT &\phantom{-}6.64 &  230&     46   &  19 &    50    \\
283 Emma          &  IN  &    -5.69       &  290&     33   &  58 &    67    \\
                  &  OUT &\phantom{-}5.66 &  628&    197   & 126 &   234    \\
569 Misa          &  IN  &    -6.23       &  319&    242   &  80 &   255    \\
                  &  OUT &\phantom{-}6.15 &  249&     85   &  62 &   105    \\
686 Gersuind      & IN   &          -3.82 & 1490&    843   & 298 & 894\\
                  & OUT  &\phantom{-}3.62 & 1436&    371   & 287 & 469\\
\hline
302 Clarissa     &  IN  &     -6.41      &    57&      11  &  14 &    18      \\
                 &  OUT &\phantom{-}6.37 &    47&       3  &  12 &    12      \\
396 Aeolia       &  IN  &     -3.09      &   100&      18  &  25 &    31    \\
                 &  OUT &\phantom{-}3.08 &    91&      15  &  23 &    27    \\
606 Brangane     &  IN  &     -3.82      &    48&       4  &  10 &    10    \\
                 &  OUT &\phantom{-}3.81 &    44&       7  &   9 &    11    \\
\hline
363 Padua         & IN   &   -5.90        & 177 &     24   &  35 &  43\\
                  & OUT  &\phantom{-}5.82 & 202 &     21   &  40 &  46\\ 
480 Hansa         &  IN  &   -3.53        &  763&    310   & 153 &  346 \\
                  &  OUT &\phantom{-}3.44 &  950&    117   & 190 &  223 \\
1303 Luthera      & IN &   -5.55          & 279 &  26 &  84 &  88 \\
                  & OUT&\phantom{-}5.52   & 273 &  34 &  82 &  89 \\
1547 Nele        &  IN  &      -3.61     &    14&       2  &   4 &  5    \\
     	         &  OUT &\phantom{-}3.61 &    15&       5  &   5 &  7    \\
\hline
9506 Telramund   & IN &        -3.55     &   205 &  56     &  41   &  71  \\
                 & OUT& \phantom{-}3.54  &   234 &  57     &  46   &  68  \\
15124 2000EZ$_{39}$&  IN  &   -6.22        &  111&     10   &  28 &   29   \\
                  &  OUT &\phantom{-}6.18 &  113&     11   &  28 &   30    \\
\hline
\end{tabular}
\end{table}

\end{document}